\documentclass[a4paper]{article}
\usepackage{lmodern}
\usepackage{amssymb,amsmath}
\usepackage{ifxetex,ifluatex}
\usepackage{fixltx2e} 
\ifnum 0\ifxetex 1\fi\ifluatex 1\fi=0 
  \usepackage[T1]{fontenc}
  \usepackage[utf8]{inputenc}
\else 
  \ifxetex
    \usepackage{mathspec}
  \else
    \usepackage{fontspec}
  \fi
  \defaultfontfeatures{Ligatures=TeX,Scale=MatchLowercase}
\fi
\IfFileExists{upquote.sty}{\usepackage{upquote}}{}
\IfFileExists{microtype.sty}{%
\usepackage{microtype}
\UseMicrotypeSet[protrusion]{basicmath} 
}{}
\usepackage[margin=1in]{geometry}
\usepackage{hyperref}
\hypersetup{unicode=true,
            pdfborder={0 0 0},
            breaklinks=true}
\usepackage{longtable,booktabs}
\usepackage{graphicx,grffile}
\makeatletter
\def\maxwidth{\ifdim\Gin@nat@width>\linewidth\linewidth\else\Gin@nat@width\fi}
\def\maxheight{\ifdim\Gin@nat@height>\textheight\textheight\else\Gin@nat@height\fi}
\makeatother
\setkeys{Gin}{width=\maxwidth,height=\maxheight,keepaspectratio}
\IfFileExists{parskip.sty}{%
\usepackage{parskip}
}{
\setlength{\parindent}{0pt}
\setlength{\parskip}{6pt plus 2pt minus 1pt}
}
\providecommand{\tightlist}{%
  \setlength{\itemsep}{0pt}\setlength{\parskip}{0pt}}
\ifx\paragraph\undefined\else
\let\oldparagraph\paragraph
\renewcommand{\paragraph}[1]{\oldparagraph{#1}\mbox{}}
\fi
\ifx\subparagraph\undefined\else
\let\oldsubparagraph\subparagraph
\renewcommand{\subparagraph}[1]{\oldsubparagraph{#1}\mbox{}}
\fi

\let\rmarkdownfootnote\footnote%
\def\footnote{\protect\rmarkdownfootnote}

\usepackage{titling}

  \title{}
  \pretitle{\vspace{\droptitle}}
  \posttitle{}
  \author{}
  \preauthor{}\postauthor{}
  \date{}
  \predate{}\postdate{}

\usepackage{amsmath}
\usepackage{amsfonts,amssymb,textcomp,gensymb}
\usepackage{bbm}

\usepackage{scrpage2}
\ifoot[]{}
\ofoot[\pagemark]{\pagemark}

\begin{document}

\section{A model-based approach to assist variety evaluation in
sunflower
crop}\label{a-model-based-approach-to-assist-variety-evaluation-in-sunflower-crop}

Pierre Casadebaig (1), Emmanuelle Mestries (2), Philippe Debaeke (1)

\begin{enumerate}
\def\labelenumi{(\arabic{enumi})}
\tightlist
\item
  AGIR, Université de Toulouse, INRA, INPT, INP-EI PURPAN,
  Castanet-Tolosan, France
\item
  Terres Inovia, Centre de recherche INRA de Toulouse, AGIR, 31326
  Castanet-Tolosan, France
\end{enumerate}

\subsection[Abstract]{\texorpdfstring{Abstract\footnote{This is the
  post-print version of the manuscript published in \emph{European
  Journal of Agronomy}
  (\href{http://dx.doi.org/10.1016/j.eja.2016.09.001}{10.1016/j.eja.2016.09.001})}}{Abstract}}\label{abstract}

Assessing the performance and the characteristics (e.g.~yield, quality,
disease resistance, abiotic stress tolerance) of new varieties is a key
component of crop performance improvement. However, the variety testing
process is presently exclusively based on experimental field approaches
which inherently reduces the number and the diversity of experienced
combinations of varieties x environmental conditions in regard of the
multiplicity of growing conditions within the cultivation area. Our aim
is to make a greater and faster use of the information issuing from
these trials using crop modeling and simulation to amplify the
environmental and agronomic conditions in which the new varieties are
tested.

In this study, we present a model-based approach to assist variety
testing and implement this approach on sunflower crop, using the SUNFLO
simulation model and a subset of 80 trials from a large
multi-environment trial (MET) conducted each year by agricultural
extension services to compare newly released sunflower hybrids. After
estimating parameter values (using plant phenotyping) to account for new
genetic material, we independently evaluated the model prediction
capacity on the MET (relative RMSE for oil yield was 16.4\%; model
accuracy was 54.4 \%) and its capacity to rank commercial hybrids for
performance level (relative RMSE was 11 \%; Kendall's \(\tau\) = 0.41, P
\textless{} 0.01). We then designed a numerical experiment by combining
the previously tested genetic and new cropping conditions (2100 virtual
trials) to determine the best varieties and related management in
representative French production regions. Finally, we proceeded to
optimize the variety-environment-management choice: growing different
varieties according to cultivation areas was a better strategy than
relying on the global adaptation of varieties. We suggest that this
approach could find operational outcomes to recommend varieties
according to environment types. Such spatial management of genetic
resources could potentially improve crop performance by reducing the
genotype-phenotype mismatch in farming environments.

\newpage

\subsection{Introduction}\label{introduction}

The purpose of plant breeding programs is to develop new crop varieties
with improved traits such as grain yield and quality across a range of
environmental conditions. As newly released crop varieties will be grown
in agro-ecologically diverse target regions, it is important to test
candidate genotypes in a representative sample of environments (climate,
soils, cropping systems) from the target region and to assess
genotype-by-environment (G \(\times\) E) interactions. G \(\times\) E
interactions correspond to significant changes in the relative
performance of the genotypes when they are evaluated in different
environments. Such interactions could be substantial in sunflower crop
as their contribution to total yield variance may range from 5 to 20 \%
(far less than the E contribution but generally more than the G
contribution) (de la Vega et al., 2001; Foucteau et al., 2001;
Marinković et al., 2011). Large and regional G \(\times\) E interactions
complicate selection for broad adaptation (de la Vega, 2012) but
identifying specifically adapted genotypes could be promoted in advisory
systems to optimize locally the variety choice (Jeuffroy et al., 2014).

New lines and varieties developed by breeders are usually subjected to
multi-environment trials (MET) to evaluate their relative performance
for a target population of environments (TPE) (DeLacy et al., 1996;
Welham et al., 2010). Today, these trials still play a key role in
variety testing during breeding programs as well as for providing
recommendations to farmers by extension services. A range of
stakeholders are involved in this experimental testing: seed companies,
extension services, cooperatives, private consultants and public
services in charge of the official registration, sharing common
experimental designs and methodological questions but having also their
own objectives and needs (Lecomte et al., 2010).

The schemes used for official variety testing of field crops are
slightly different from one country to another. However, there is a
common feature in Europe (Van Waes, 2009) where official variety trials
are arranged by dedicated institutes for the examination of value for
cultivation and use (VCU). These VCU trials aim to emulate practical
cultivation conditions in areas suitable for each crop. Before its
commercial release, each variety undergoes VCU trials for at least two
years before entering in the National List of Plant Varieties.
Thereafter the variety undergoes trials for another 2-3 years to
identify the best performing new varieties under local conditions and
provide descriptions of their main agronomic and quality
characteristics. These trials result in the production of Recommended
Variety Lists. The recommendations for variety cultivation are published
either as national summaries or as regional bulletins. Altogether the
testing of a variety in official variety trials takes 3 to 6 years
depending on the countries and crops.

In France, a new variety is registered in the Official Catalog of Plant
Varieties after two or three years of successful field testing
(pre-registration) conducted by GEVES\footnote{Groupe d'Etude et de
  Contrôle des Variétés et des Semences} (in charge of official seed and
variety testing) using METs. Then newly released varieties are tested
over a wider area and with more trials to determine their regional
performances by technical institutes in charge of extension services
(post-registration). Regarding sunflower, the French technical institute
for oilseed crops (\emph{Terres Inovia}\footnote{Institut technique des
  oléagineux, des protéagineux et du chanvre industriel (formerly
  \emph{CETIOM})}) is in charge of the post-registration activity. In
parallel, seed companies and cooperatives also conduct private testing
to enhance their own expertise.

However, these METs are conducted at great expense and there is
potential to make greater and faster use of the information issuing from
these numerous trials for registration and further recommendation.
Focusing on sunflower, we identified several weaknesses of the current
VCU design exclusively based on field testing in France (Debaeke et al.,
2010, 2011):

\begin{enumerate}
\def\labelenumi{(\arabic{enumi})}
\item
  Pre- and post-registration trials (respectively conducted by
  \emph{GEVES} and \emph{Terres Inovia}) are not pooled for a common
  analysis and the number of trials in each MET steadily decreases with
  time. Currently, each sunflower variety is thus assessed on 15 to 45
  trials then the results are clustered in one to four regional pools
  for simplifying recommendations. When pooling all the maturity groups
  and oil composition types (linoleic vs oleic), about a hundred trials
  are carried out each year to assess the agronomical, technological and
  environmental value of newly released varieties. Consequently, the
  independant analysis of networks (and years) increase the risk of
  matching the wrong management or location to a variety, and also mean
  that a variety of value to a particular location might be discarded
  prematurely. Additionaly, the crop management and soil conditions
  observed in evaluation networks does not match farmer's conditions
  where a greater proportion of shallow soils and low-input management
  conditions (no irrigation, low plant density) were observed (Debaeke
  et al., 2012). Consequently, the capacity of these METs to represent
  the target population of environments is decreasing from year to year,
  because of the sheer decrease of the number of trials, and because of
  their design. Although this experimental network covers the main
  regions of sunflower production in France, more diverse environmental
  conditions (soil, weather, crop management) would undoubtedly improve
  the assessment of yield stability.
\item
  In many cases, if yields are low at particular locations because of
  drought stress, the entire trial will be rejected because of increased
  error variances. We argue that when these data are discarded for
  statistical reasons, valuable information is lost. More importantly,
  released varieties can be biased towards those that perform well under
  ideal conditions, but run the risk of performing poorly when water is
  limiting (Pidgeon et al., 2006).
\item
  Only a few criteria are used to assess the performance of new
  sunflower varieties and they are restricted to final productivity,
  grain quality, earliness and tolerance to major diseases (Table 1).
  Although sunflower is a summer crop, grown without irrigation in
  shallow to moderately-deep soils, no routine evaluation of drought
  tolerance traits is performed (except earliness at anthesis which is a
  drought escaping trait). Therefore, current evaluation criteria
  underestimate the rusticity of some varieties, eventually performing
  better under water deficit. Moreover, there has been little, if any,
  sound characterization of the physical environment (e.g.~available
  soil water) and of the constraints perceived by plants (water and
  nitrogen stresses, disease severity, \ldots{}). A proper
  characterization of the environments over the MET (e.g. Chenu et al.,
  2011) would facilitate the analysis of G \(\times\) E interactions and
  the clustering of trials having similar stress patterns.
\item
  On each site, a single crop management system is tested, independently
  of variety-specific requirements, which impedes the proposal of a
  specific ``variety-management'' recommendation per type of
  environment. Therefore, \href{http://www.myvar.fr/}{MyVar}, a decision
  support tool developed by \emph{Terres Inovia} in 2014, does not
  provide cultivar recommendations for crop management but only
  indicates characteristics of suitable varieties.
\end{enumerate}

\begin{longtable}[]{@{}cccc@{}}
\toprule
\begin{minipage}[b]{0.24\columnwidth}\centering\strut
Criteria\strut
\end{minipage} & \begin{minipage}[b]{0.28\columnwidth}\centering\strut
Entry\strut
\end{minipage} & \begin{minipage}[b]{0.09\columnwidth}\centering\strut
Levels\strut
\end{minipage} & \begin{minipage}[b]{0.29\columnwidth}\centering\strut
Details\strut
\end{minipage}\tabularnewline
\midrule
\endhead
\begin{minipage}[t]{0.24\columnwidth}\centering\strut
General information\strut
\end{minipage} & \begin{minipage}[t]{0.28\columnwidth}\centering\strut
Breeding company\strut
\end{minipage} & \begin{minipage}[t]{0.09\columnwidth}\centering\strut
-\strut
\end{minipage} & \begin{minipage}[t]{0.29\columnwidth}\centering\strut
\strut
\end{minipage}\tabularnewline
\begin{minipage}[t]{0.24\columnwidth}\centering\strut
\strut
\end{minipage} & \begin{minipage}[t]{0.28\columnwidth}\centering\strut
Year of release\strut
\end{minipage} & \begin{minipage}[t]{0.09\columnwidth}\centering\strut
3\strut
\end{minipage} & \begin{minipage}[t]{0.29\columnwidth}\centering\strut
\textless{}2005, 2005-2010, \textgreater{}2010\strut
\end{minipage}\tabularnewline
\begin{minipage}[t]{0.24\columnwidth}\centering\strut
\strut
\end{minipage} & \begin{minipage}[t]{0.28\columnwidth}\centering\strut
Registration EU Country\strut
\end{minipage} & \begin{minipage}[t]{0.09\columnwidth}\centering\strut
2\strut
\end{minipage} & \begin{minipage}[t]{0.29\columnwidth}\centering\strut
France, Abroad\strut
\end{minipage}\tabularnewline
\begin{minipage}[t]{0.24\columnwidth}\centering\strut
Plant phenology\strut
\end{minipage} & \begin{minipage}[t]{0.28\columnwidth}\centering\strut
Anthesis earliness\strut
\end{minipage} & \begin{minipage}[t]{0.09\columnwidth}\centering\strut
5\strut
\end{minipage} & \begin{minipage}[t]{0.29\columnwidth}\centering\strut
very early to late\strut
\end{minipage}\tabularnewline
\begin{minipage}[t]{0.24\columnwidth}\centering\strut
\strut
\end{minipage} & \begin{minipage}[t]{0.28\columnwidth}\centering\strut
Maturity earliness\strut
\end{minipage} & \begin{minipage}[t]{0.09\columnwidth}\centering\strut
5\strut
\end{minipage} & \begin{minipage}[t]{0.29\columnwidth}\centering\strut
very early to late\strut
\end{minipage}\tabularnewline
\begin{minipage}[t]{0.24\columnwidth}\centering\strut
Plant morphology\strut
\end{minipage} & \begin{minipage}[t]{0.28\columnwidth}\centering\strut
Plant height\strut
\end{minipage} & \begin{minipage}[t]{0.09\columnwidth}\centering\strut
3\strut
\end{minipage} & \begin{minipage}[t]{0.29\columnwidth}\centering\strut
short, medium, tall\strut
\end{minipage}\tabularnewline
\begin{minipage}[t]{0.24\columnwidth}\centering\strut
Disease tolerance\strut
\end{minipage} & \begin{minipage}[t]{0.28\columnwidth}\centering\strut
Phomopsis stem canker\strut
\end{minipage} & \begin{minipage}[t]{0.09\columnwidth}\centering\strut
5\strut
\end{minipage} & \begin{minipage}[t]{0.29\columnwidth}\centering\strut
\strut
\end{minipage}\tabularnewline
\begin{minipage}[t]{0.24\columnwidth}\centering\strut
\strut
\end{minipage} & \begin{minipage}[t]{0.28\columnwidth}\centering\strut
Sclerotinia head rot\strut
\end{minipage} & \begin{minipage}[t]{0.09\columnwidth}\centering\strut
4\strut
\end{minipage} & \begin{minipage}[t]{0.29\columnwidth}\centering\strut
\strut
\end{minipage}\tabularnewline
\begin{minipage}[t]{0.24\columnwidth}\centering\strut
\strut
\end{minipage} & \begin{minipage}[t]{0.28\columnwidth}\centering\strut
Sclerotinia basal stalk rot\strut
\end{minipage} & \begin{minipage}[t]{0.09\columnwidth}\centering\strut
3\strut
\end{minipage} & \begin{minipage}[t]{0.29\columnwidth}\centering\strut
\strut
\end{minipage}\tabularnewline
\begin{minipage}[t]{0.24\columnwidth}\centering\strut
\strut
\end{minipage} & \begin{minipage}[t]{0.28\columnwidth}\centering\strut
Verticillium wilt\strut
\end{minipage} & \begin{minipage}[t]{0.09\columnwidth}\centering\strut
4\strut
\end{minipage} & \begin{minipage}[t]{0.29\columnwidth}\centering\strut
\strut
\end{minipage}\tabularnewline
\begin{minipage}[t]{0.24\columnwidth}\centering\strut
Downy mildew resistance\strut
\end{minipage} & \begin{minipage}[t]{0.28\columnwidth}\centering\strut
Resistance profile\strut
\end{minipage} & \begin{minipage}[t]{0.09\columnwidth}\centering\strut
3\strut
\end{minipage} & \begin{minipage}[t]{0.29\columnwidth}\centering\strut
RM9, RM8, other RMs\strut
\end{minipage}\tabularnewline
\begin{minipage}[t]{0.24\columnwidth}\centering\strut
Herbicide tolerance\strut
\end{minipage} & \begin{minipage}[t]{0.28\columnwidth}\centering\strut
Technology employed\strut
\end{minipage} & \begin{minipage}[t]{0.09\columnwidth}\centering\strut
3\strut
\end{minipage} & \begin{minipage}[t]{0.29\columnwidth}\centering\strut
none, Clearfield, Express Sun\strut
\end{minipage}\tabularnewline
\begin{minipage}[t]{0.24\columnwidth}\centering\strut
Seed characteristics\strut
\end{minipage} & \begin{minipage}[t]{0.28\columnwidth}\centering\strut
Thousand seed weight\strut
\end{minipage} & \begin{minipage}[t]{0.09\columnwidth}\centering\strut
3\strut
\end{minipage} & \begin{minipage}[t]{0.29\columnwidth}\centering\strut
low, medium, high\strut
\end{minipage}\tabularnewline
\begin{minipage}[t]{0.24\columnwidth}\centering\strut
Oil characteristics\strut
\end{minipage} & \begin{minipage}[t]{0.28\columnwidth}\centering\strut
Oil concentration\strut
\end{minipage} & \begin{minipage}[t]{0.09\columnwidth}\centering\strut
4\strut
\end{minipage} & \begin{minipage}[t]{0.29\columnwidth}\centering\strut
Low, medium, high, very high\strut
\end{minipage}\tabularnewline
\begin{minipage}[t]{0.24\columnwidth}\centering\strut
\strut
\end{minipage} & \begin{minipage}[t]{0.28\columnwidth}\centering\strut
Oil quality\strut
\end{minipage} & \begin{minipage}[t]{0.09\columnwidth}\centering\strut
2\strut
\end{minipage} & \begin{minipage}[t]{0.29\columnwidth}\centering\strut
high oleic, linoleic (mid-oleic)\strut
\end{minipage}\tabularnewline
\begin{minipage}[t]{0.24\columnwidth}\centering\strut
Yield performance\strut
\end{minipage} & \begin{minipage}[t]{0.28\columnwidth}\centering\strut
Performance level\strut
\end{minipage} & \begin{minipage}[t]{0.09\columnwidth}\centering\strut
5\strut
\end{minipage} & \begin{minipage}[t]{0.29\columnwidth}\centering\strut
Scale depending on multi-location field trials\strut
\end{minipage}\tabularnewline
\bottomrule
\end{longtable}

\begin{quote}
\textbf{Table 1. Criteria available in France to choose a sunflower
variety.} These criteria are included in the
\href{http://www.myvar.fr}{MyVar} web tool developed by Terres Inovia.
\end{quote}

While all these stated problems increase the risk of matching the wrong
management or location to a variety, they may also mean that a variety
of value to a particular location is discarded prematurely.
Consequently, statistical analysis of the data collected on METs has
received a lot of attention, largely in response to the difficulties
caused by G \(\times\) E interactions (Malosetti et al., 2013; e.g.
Piepho et al., 2012). Besides, in spite of their potential interest,
dynamic crop models have not been used extensively to explain and
predict G \(\times\) E interactions (Bustos-Korts et al., 2016; Chapman,
2008; Chapman et al., 2002). Generally speaking, we may consider that
environmental characterization and diagnosis of yield limiting factors
are not sufficiently practiced by breeders and advisory services to
exploit the G \(\times\) E interactions that could be detected. We
assume that crop modeling and simulation could significantly improve the
efficacy of this experimental assessment by its ability to explore
untested conditions and by giving access to soil and plant variables
that are not measured in variety trials. For instance, the SUNFLO crop
model (Casadebaig et al., 2011; Lecoeur et al., 2011) was developed to
simulate on a daily step the response of sunflower genotypes to various
soil-weather environments and management options (sowing date, plant
density, nitrogen fertilization, irrigation) and some applications in
variety testing and plant breeding have been suggested (Casadebaig and
Debaeke, 2011; Casadebaig et al., 2014; Jeuffroy et al., 2014).

In this study, we will explore how field-based assessment can leverage
simulation, either to characterise environments (simpler, E problem) or
to rank cultivars (harder, G \(\times\) E problem). The objective of
this contribution is to develop an integrated framework for variety
evaluation of sunflower based on crop modeling in order to widen and
complete the current information on sunflower varieties provided by
official advisory, extension services or private seed companies for a
range of environmental and agronomic conditions. The potential use of
this framework will be illustrated as a proof of concept.

\subsection{A framework to include crop modeling in the current variety
evaluation
process}\label{a-framework-to-include-crop-modeling-in-the-current-variety-evaluation-process}

A model-based approach was designed to assist variety evaluation in due
time and to amplify the environmental and agronomic conditions in which
the varieties are routinely tested. Four steps were identified and
integrated in the current process of pre- and post-registration used in
France by \emph{GEVES} and \emph{Terres Inovia}, targeting the extension
services and seed companies to define the proper use of newly released
varieties (Figure 1). For instance questions such as \emph{In which
pedo-climatic area should I promote this material? In association to
which crop management ?} are addressed by this model-based approach.

\includegraphics{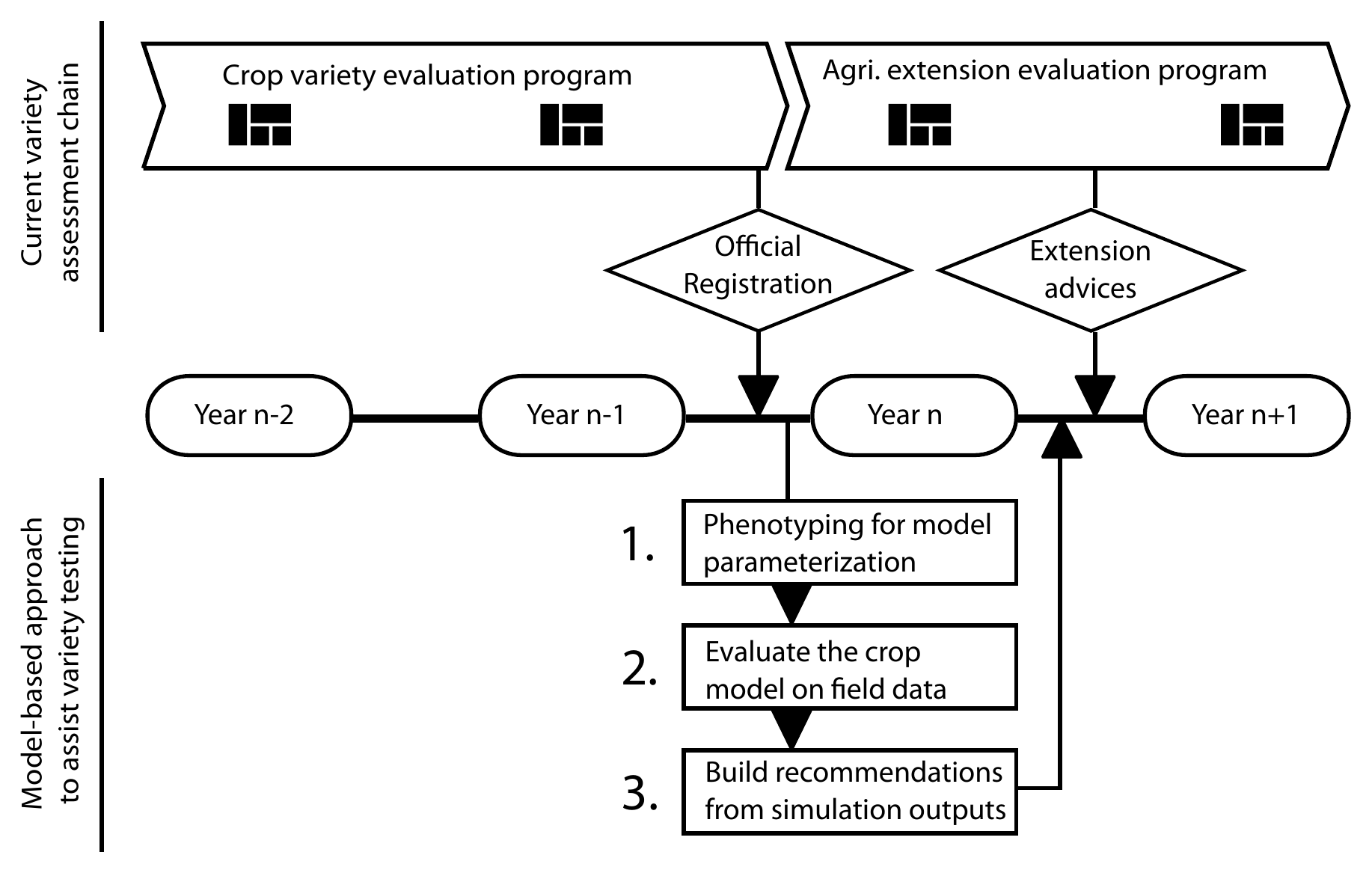}

\begin{quote}
\textbf{Figure 1. A framework to include crop modeling in the variety
evaluation process.} The representation of the variety evaluation chain
(upper part in the diagram) is based on the current French system, with
two years on trials before variety release (variety evaluation MET), and
two year after (agricultural extension MET). The presented approach
leverage existing trials to get informations on tested varieties and to
evaluate the crop model (steps 1-2). After designing numerical
experiments, simulation would then provides recommendation for variety
choice, accounting for the diversity of growing conditions and climatic
uncertainty (steps 3).
\end{quote}

\textbf{Step 1. Phenotype newly released varieties to estimate parameter
values.} Phenotyping will be limited to newly released varieties,
distributed for commercial development. Each year, about 15-25 sunflower
varieties are registered by CTPS\footnote{Comité Technique Permanent de
  la Sélection des plantes cultivées} in France. Although the varieties
have been already tested during two years by \emph{GEVES} before their
official registration, they will be tested during at least one more year
over a larger MET network by \emph{Terres Inovia} to evaluate their
regional adaptation (Mestries and Jouffret, 2002). The usual phenotyping
which is currently performed for variety evaluation is limited to five
agronomic variables: anthesis date, plant height, achene moisture at
harvest, grain yield and oil concentration (CTPS, 2014). In two selected
field locations, newly released varieties will be examined in microplots
to measure the 10 out of 12 of the genotype-dependent parameters of the
SUNFLO crop model (Table 2). These specific trials target data on
phenology, architecture and yield build-up. The response of leaf
expansion and transpiration to soil water deficit (2 additional
parameters) will be determined in controlled conditions (Casadebaig et
al., 2008; Lecoeur et al., 2011).

\textbf{Step 2. Evaluate the SUNFLO crop model using variety evaluation
networks.} Once parameterized to represent new varieties, the SUNFLO
model will be evaluated on the \emph{GEVES} and \emph{Terres Inovia}
networks to determine its predictive quality for this new genetic
material. Depending on its performance, the model will be stated as
valid or not for the range of commercial varieties. If valid, the model
will be used to run numerical experiments in next step.

\textbf{Step 3. Design numerical experiments and provide variety
recommendation.} Stakeholders will design numerical experiments
combining varieties, target environments and management options. The
SUNFLO model will then be run using 30 years of meteorological data to
cope with climatic uncertainty. Simulated rankings (mean and standard
deviation) will be produced on grain and oil yield for each combination
of soil, climate and management. The previous computational results may
be completed with other datasets not available through simulation, such
as the tolerance of the varieties to abiotic factors (e.g.~fungal
diseases from Table 1). Advisers could choose the best varieties and the
management to adapt and disseminate their recommendations through usual
media (publications, web, meetings, \ldots{}). In this study, this final
step will not be illustrated.

\subsection{Materials and Methods}\label{materials-and-methods}

\subsubsection{The SUNFLO crop model}\label{the-sunflo-crop-model}

The SUNFLO model is the core of the approach. SUNFLO is a process-based
model for the sunflower crop which was developed to simulate the grain
yield and oil concentration as a function of time, environment (soil and
climate), management practices (irrigation, fertilization, crop density)
and genetic diversity, through genotype-dependent parameters (Casadebaig
et al., 2011; Debaeke et al., 2010; Lecoeur et al., 2011) (Figure 2,
Table 2.). The model simulates the main soil and plant processes: root
growth, soil water and nitrogen content, plant transpiration and
nitrogen uptake, leaf expansion and senescence and biomass accumulation,
as a function of main environmental constraints (temperature, radiation,
water and nitrogen deficit).

This model is based on a conceptual framework initially proposed by
Monteith (1977) and now shared by a large family of crop models (Brisson
et al., 2003; Holzworth et al., 2014; Jones et al., 2003). In this
framework, the daily crop dry biomass (\(DM_t\)) is calculated as an
ordinary difference equation (eq. \ref{eq:ode}) function of incident
photosynthetically active radiation (\(PAR\), MJ m\textsuperscript{-2}),
light interception efficiency (\(1-exp^{-k \cdot LAI}\)) and radiation
use efficiency (\(RUE\), g MJ\textsuperscript{-1}, Monteith (1994)). The
light interception efficiency is based on Beer-Lambert's law as a
function of leaf area index (\(LAI\)) and light extinction coefficient
(\(k\)). The SUNFLO model is based on a distributed approach of leaf
expansion and senescence rather than a homogeneous canopy layer
(\emph{big leaf}) and intercepted radiation per LAI is used to to drive
leaf expansion response to cropping density (Rey et al., 2008).

\begin{equation}
  DM_t = DM_{t-1} + RUE_t \cdot (1-exp^{-k \cdot LAI_t}) \cdot PAR_t
  \label{eq:ode}
\end{equation}

Thus, the simulated G \(\times\) E interactions result from the impact
of genotype-dependent traits (phenology, architecture, biomass
allocation) on the capture of environmental resources (radiation, water,
nitrogen) and on the differential responses of the genotypes to
environmental constraints in a dynamic feed-back. The model divides the
crop cycle into 6 phenological phases using thermal time (base 4.8 °C,
Granier and Tardieu (1998)): (1) sowing (A0) - emergence (A2), (2)
emergence - floral initiation (E1), (3) floral initiation - early
anthesis (F1), (4) early anthesis - early grain-filling (M0), (5) early
to late grain filling (M3, physiological maturity) and (6) physiological
maturity to harvest time (M4). Each phenological stage induces
differential physiological processes. Water and nutrition uptake are
simulated daily and computed stress variables impact crop phenology,
plant transpiration, leaf expansion, and biomass accumulation. Yield is
estimated through harvest index (HI, seed:aerial biomass ratio) rather
than from yield components. Harvest index and oil concentration values
at harvest are estimated using multiple linear regressions using two
type of predictors: (1) simulated variables estimated in the
process-based part of the model (e.g.~sum of intercepted light, nitrogen
and water deficit) and genotype-dependent parameters estimated in
non-limiting field conditions (potential harvest index and oil content)
from METs.

\includegraphics[width=0.7\textwidth]{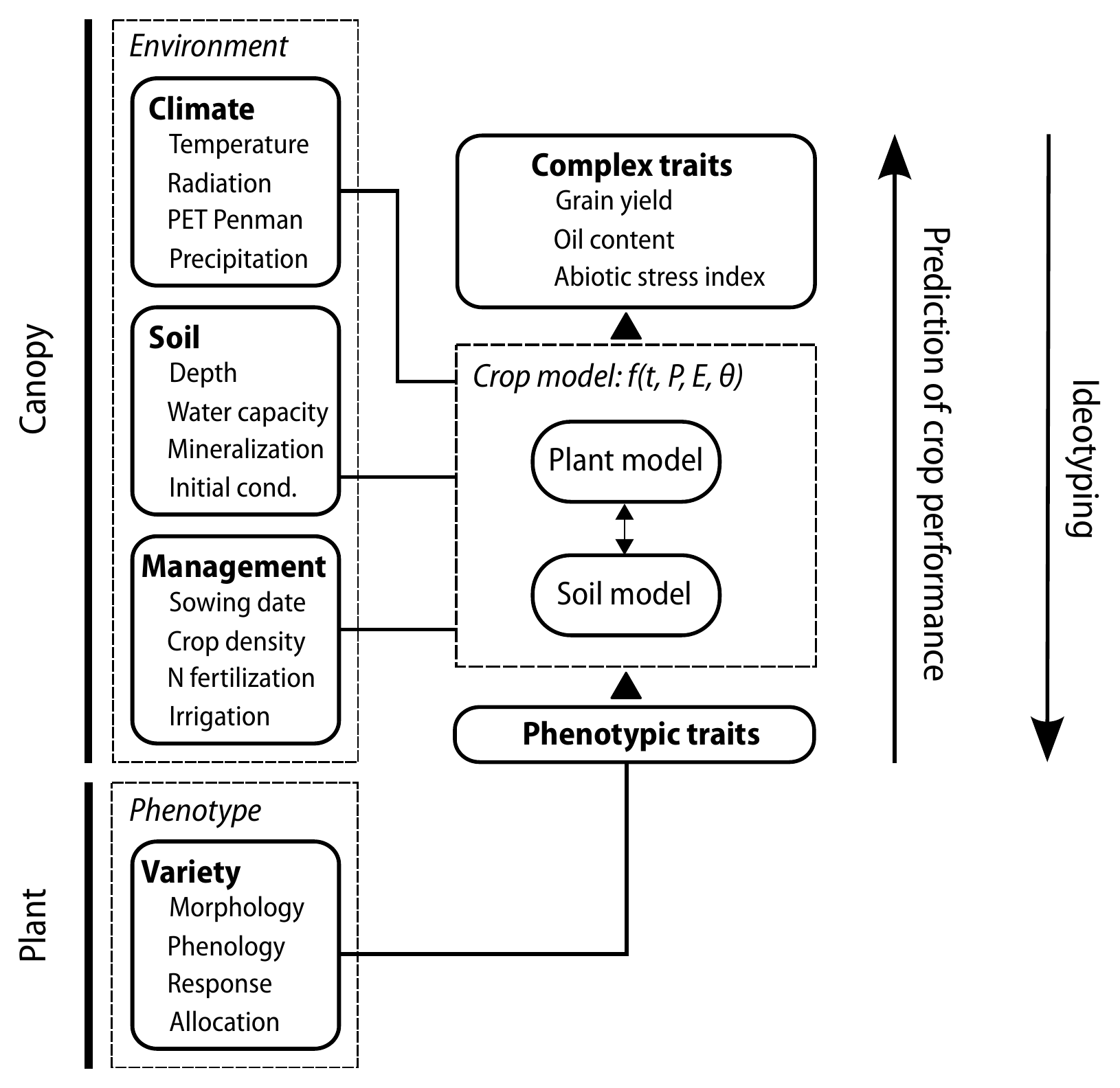}

\begin{quote}
\textbf{Figure 2. Schematic representation of the SUNFLO crop model.}
Soil-Plant system is described as a function of time, environmental
inputs and phenotypic inputs (left column). Phenotypic traits are used
as genotype-dependent parameters in the model.
\end{quote}

Each variety is currently described by 12 parameters, i.e.~phenotypic
traits measurable in field and controlled conditions for newly released
material (Casadebaig et al., 2008, 2014; Debaeke et al., 2010). The
parameters can be sorted in four groups: phenology (4 parameters), leaf
architecture (4), response to water constraint (2), and biomass
allocation to the grains (2) (Table 2). Furthermore, we assumed that
this set of 12 traits were sufficient to describe the adaptation of a
variety to a location. Other traits, not accounted for in the crop model
can also drive plant response to environment (such limitations are
further discussed in Casadebaig et al., 2011).

\footnotesize

\begin{longtable}[]{@{}llllccccrrr@{}}
\toprule
\begin{minipage}[b]{0.08\columnwidth}\raggedright\strut
Process\strut
\end{minipage} & \begin{minipage}[b]{0.05\columnwidth}\raggedright\strut
Symbol\strut
\end{minipage} & \begin{minipage}[b]{0.19\columnwidth}\raggedright\strut
Function\strut
\end{minipage} & \begin{minipage}[b]{0.06\columnwidth}\raggedright\strut
Unit\strut
\end{minipage} & \begin{minipage}[b]{0.06\columnwidth}\centering\strut
Field\strut
\end{minipage} & \begin{minipage}[b]{0.08\columnwidth}\centering\strut
Controlled\strut
\end{minipage} & \begin{minipage}[b]{0.05\columnwidth}\centering\strut
Model\strut
\end{minipage} & \begin{minipage}[b]{0.04\columnwidth}\centering\strut
MET\strut
\end{minipage} & \begin{minipage}[b]{0.04\columnwidth}\raggedleft\strut
mean\strut
\end{minipage} & \begin{minipage}[b]{0.04\columnwidth}\raggedleft\strut
min\strut
\end{minipage} & \begin{minipage}[b]{0.04\columnwidth}\raggedleft\strut
max\strut
\end{minipage}\tabularnewline
\midrule
\endhead
\begin{minipage}[t]{0.08\columnwidth}\raggedright\strut
Phenology\strut
\end{minipage} & \begin{minipage}[t]{0.05\columnwidth}\raggedright\strut
TDE1\strut
\end{minipage} & \begin{minipage}[t]{0.19\columnwidth}\raggedright\strut
Temperature sum to floral initiation\strut
\end{minipage} & \begin{minipage}[t]{0.06\columnwidth}\raggedright\strut
\(C.d\)\strut
\end{minipage} & \begin{minipage}[t]{0.06\columnwidth}\centering\strut
Possible\strut
\end{minipage} & \begin{minipage}[t]{0.08\columnwidth}\centering\strut
No\strut
\end{minipage} & \begin{minipage}[t]{0.05\columnwidth}\centering\strut
Yes\strut
\end{minipage} & \begin{minipage}[t]{0.04\columnwidth}\centering\strut
No\strut
\end{minipage} & \begin{minipage}[t]{0.04\columnwidth}\raggedleft\strut
482\strut
\end{minipage} & \begin{minipage}[t]{0.04\columnwidth}\raggedleft\strut
441\strut
\end{minipage} & \begin{minipage}[t]{0.04\columnwidth}\raggedleft\strut
523\strut
\end{minipage}\tabularnewline
\begin{minipage}[t]{0.08\columnwidth}\raggedright\strut
Phenology\strut
\end{minipage} & \begin{minipage}[t]{0.05\columnwidth}\raggedright\strut
TDF1\strut
\end{minipage} & \begin{minipage}[t]{0.19\columnwidth}\raggedright\strut
Temperature sum from emergence to the beginning of flowering\strut
\end{minipage} & \begin{minipage}[t]{0.06\columnwidth}\raggedright\strut
\(C.d\)\strut
\end{minipage} & \begin{minipage}[t]{0.06\columnwidth}\centering\strut
Yes\strut
\end{minipage} & \begin{minipage}[t]{0.08\columnwidth}\centering\strut
No\strut
\end{minipage} & \begin{minipage}[t]{0.05\columnwidth}\centering\strut
No\strut
\end{minipage} & \begin{minipage}[t]{0.04\columnwidth}\centering\strut
Yes\strut
\end{minipage} & \begin{minipage}[t]{0.04\columnwidth}\raggedleft\strut
836\strut
\end{minipage} & \begin{minipage}[t]{0.04\columnwidth}\raggedleft\strut
765\strut
\end{minipage} & \begin{minipage}[t]{0.04\columnwidth}\raggedleft\strut
907\strut
\end{minipage}\tabularnewline
\begin{minipage}[t]{0.08\columnwidth}\raggedright\strut
Phenology\strut
\end{minipage} & \begin{minipage}[t]{0.05\columnwidth}\raggedright\strut
TDM0\strut
\end{minipage} & \begin{minipage}[t]{0.19\columnwidth}\raggedright\strut
Temperature sum from emergence to the beginning of grain filling\strut
\end{minipage} & \begin{minipage}[t]{0.06\columnwidth}\raggedright\strut
\(C.d\)\strut
\end{minipage} & \begin{minipage}[t]{0.06\columnwidth}\centering\strut
Possible\strut
\end{minipage} & \begin{minipage}[t]{0.08\columnwidth}\centering\strut
No\strut
\end{minipage} & \begin{minipage}[t]{0.05\columnwidth}\centering\strut
Yes\strut
\end{minipage} & \begin{minipage}[t]{0.04\columnwidth}\centering\strut
No\strut
\end{minipage} & \begin{minipage}[t]{0.04\columnwidth}\raggedleft\strut
1083\strut
\end{minipage} & \begin{minipage}[t]{0.04\columnwidth}\raggedleft\strut
1012\strut
\end{minipage} & \begin{minipage}[t]{0.04\columnwidth}\raggedleft\strut
1154\strut
\end{minipage}\tabularnewline
\begin{minipage}[t]{0.08\columnwidth}\raggedright\strut
Phenology\strut
\end{minipage} & \begin{minipage}[t]{0.05\columnwidth}\raggedright\strut
TDM3\strut
\end{minipage} & \begin{minipage}[t]{0.19\columnwidth}\raggedright\strut
Temperature sum from emergence to seed physiological maturity\strut
\end{minipage} & \begin{minipage}[t]{0.06\columnwidth}\raggedright\strut
\(C.d\)\strut
\end{minipage} & \begin{minipage}[t]{0.06\columnwidth}\centering\strut
Yes\strut
\end{minipage} & \begin{minipage}[t]{0.08\columnwidth}\centering\strut
No\strut
\end{minipage} & \begin{minipage}[t]{0.05\columnwidth}\centering\strut
No\strut
\end{minipage} & \begin{minipage}[t]{0.04\columnwidth}\centering\strut
No\strut
\end{minipage} & \begin{minipage}[t]{0.04\columnwidth}\raggedleft\strut
1673\strut
\end{minipage} & \begin{minipage}[t]{0.04\columnwidth}\raggedleft\strut
1538\strut
\end{minipage} & \begin{minipage}[t]{0.04\columnwidth}\raggedleft\strut
1831\strut
\end{minipage}\tabularnewline
\begin{minipage}[t]{0.08\columnwidth}\raggedright\strut
Architecture\strut
\end{minipage} & \begin{minipage}[t]{0.05\columnwidth}\raggedright\strut
TLN\strut
\end{minipage} & \begin{minipage}[t]{0.19\columnwidth}\raggedright\strut
Potential number of leaves at flowering\strut
\end{minipage} & \begin{minipage}[t]{0.06\columnwidth}\raggedright\strut
\(leaf\)\strut
\end{minipage} & \begin{minipage}[t]{0.06\columnwidth}\centering\strut
Yes\strut
\end{minipage} & \begin{minipage}[t]{0.08\columnwidth}\centering\strut
Possible\strut
\end{minipage} & \begin{minipage}[t]{0.05\columnwidth}\centering\strut
No\strut
\end{minipage} & \begin{minipage}[t]{0.04\columnwidth}\centering\strut
No\strut
\end{minipage} & \begin{minipage}[t]{0.04\columnwidth}\raggedleft\strut
29\strut
\end{minipage} & \begin{minipage}[t]{0.04\columnwidth}\raggedleft\strut
22\strut
\end{minipage} & \begin{minipage}[t]{0.04\columnwidth}\raggedleft\strut
37\strut
\end{minipage}\tabularnewline
\begin{minipage}[t]{0.08\columnwidth}\raggedright\strut
Architecture\strut
\end{minipage} & \begin{minipage}[t]{0.05\columnwidth}\raggedright\strut
LLH\strut
\end{minipage} & \begin{minipage}[t]{0.19\columnwidth}\raggedright\strut
Potential rank of the plant largest leaf at flowering\strut
\end{minipage} & \begin{minipage}[t]{0.06\columnwidth}\raggedright\strut
\(leaf\)\strut
\end{minipage} & \begin{minipage}[t]{0.06\columnwidth}\centering\strut
Yes\strut
\end{minipage} & \begin{minipage}[t]{0.08\columnwidth}\centering\strut
Possible\strut
\end{minipage} & \begin{minipage}[t]{0.05\columnwidth}\centering\strut
No\strut
\end{minipage} & \begin{minipage}[t]{0.04\columnwidth}\centering\strut
No\strut
\end{minipage} & \begin{minipage}[t]{0.04\columnwidth}\raggedleft\strut
17\strut
\end{minipage} & \begin{minipage}[t]{0.04\columnwidth}\raggedleft\strut
13\strut
\end{minipage} & \begin{minipage}[t]{0.04\columnwidth}\raggedleft\strut
21\strut
\end{minipage}\tabularnewline
\begin{minipage}[t]{0.08\columnwidth}\raggedright\strut
Architecture\strut
\end{minipage} & \begin{minipage}[t]{0.05\columnwidth}\raggedright\strut
LLS\strut
\end{minipage} & \begin{minipage}[t]{0.19\columnwidth}\raggedright\strut
Potential area of the plant largest leaf at flowering\strut
\end{minipage} & \begin{minipage}[t]{0.06\columnwidth}\raggedright\strut
\(cm^2\)\strut
\end{minipage} & \begin{minipage}[t]{0.06\columnwidth}\centering\strut
Yes\strut
\end{minipage} & \begin{minipage}[t]{0.08\columnwidth}\centering\strut
Possible\strut
\end{minipage} & \begin{minipage}[t]{0.05\columnwidth}\centering\strut
No\strut
\end{minipage} & \begin{minipage}[t]{0.04\columnwidth}\centering\strut
No\strut
\end{minipage} & \begin{minipage}[t]{0.04\columnwidth}\raggedleft\strut
450\strut
\end{minipage} & \begin{minipage}[t]{0.04\columnwidth}\raggedleft\strut
334\strut
\end{minipage} & \begin{minipage}[t]{0.04\columnwidth}\raggedleft\strut
670\strut
\end{minipage}\tabularnewline
\begin{minipage}[t]{0.08\columnwidth}\raggedright\strut
Architecture\strut
\end{minipage} & \begin{minipage}[t]{0.05\columnwidth}\raggedright\strut
K\strut
\end{minipage} & \begin{minipage}[t]{0.19\columnwidth}\raggedright\strut
Light extinction coefficient during vegetative growth\strut
\end{minipage} & \begin{minipage}[t]{0.06\columnwidth}\raggedright\strut
-\strut
\end{minipage} & \begin{minipage}[t]{0.06\columnwidth}\centering\strut
Difficult\strut
\end{minipage} & \begin{minipage}[t]{0.08\columnwidth}\centering\strut
No\strut
\end{minipage} & \begin{minipage}[t]{0.05\columnwidth}\centering\strut
Yes\strut
\end{minipage} & \begin{minipage}[t]{0.04\columnwidth}\centering\strut
No\strut
\end{minipage} & \begin{minipage}[t]{0.04\columnwidth}\raggedleft\strut
0.89\strut
\end{minipage} & \begin{minipage}[t]{0.04\columnwidth}\raggedleft\strut
0.78\strut
\end{minipage} & \begin{minipage}[t]{0.04\columnwidth}\raggedleft\strut
0.95\strut
\end{minipage}\tabularnewline
\begin{minipage}[t]{0.08\columnwidth}\raggedright\strut
Response\strut
\end{minipage} & \begin{minipage}[t]{0.05\columnwidth}\raggedright\strut
LE\strut
\end{minipage} & \begin{minipage}[t]{0.19\columnwidth}\raggedright\strut
Threshold for leaf expansion response to water stress\strut
\end{minipage} & \begin{minipage}[t]{0.06\columnwidth}\raggedright\strut
-\strut
\end{minipage} & \begin{minipage}[t]{0.06\columnwidth}\centering\strut
No\strut
\end{minipage} & \begin{minipage}[t]{0.08\columnwidth}\centering\strut
Yes\strut
\end{minipage} & \begin{minipage}[t]{0.05\columnwidth}\centering\strut
No\strut
\end{minipage} & \begin{minipage}[t]{0.04\columnwidth}\centering\strut
No\strut
\end{minipage} & \begin{minipage}[t]{0.04\columnwidth}\raggedleft\strut
-4.4\strut
\end{minipage} & \begin{minipage}[t]{0.04\columnwidth}\raggedleft\strut
-16\strut
\end{minipage} & \begin{minipage}[t]{0.04\columnwidth}\raggedleft\strut
-2.3\strut
\end{minipage}\tabularnewline
\begin{minipage}[t]{0.08\columnwidth}\raggedright\strut
Response\strut
\end{minipage} & \begin{minipage}[t]{0.05\columnwidth}\raggedright\strut
TR\strut
\end{minipage} & \begin{minipage}[t]{0.19\columnwidth}\raggedright\strut
Threshold for stomatal conductance response to water stress\strut
\end{minipage} & \begin{minipage}[t]{0.06\columnwidth}\raggedright\strut
-\strut
\end{minipage} & \begin{minipage}[t]{0.06\columnwidth}\centering\strut
No\strut
\end{minipage} & \begin{minipage}[t]{0.08\columnwidth}\centering\strut
Yes\strut
\end{minipage} & \begin{minipage}[t]{0.05\columnwidth}\centering\strut
No\strut
\end{minipage} & \begin{minipage}[t]{0.04\columnwidth}\centering\strut
No\strut
\end{minipage} & \begin{minipage}[t]{0.04\columnwidth}\raggedleft\strut
-9.8\strut
\end{minipage} & \begin{minipage}[t]{0.04\columnwidth}\raggedleft\strut
-14\strut
\end{minipage} & \begin{minipage}[t]{0.04\columnwidth}\raggedleft\strut
-5.8\strut
\end{minipage}\tabularnewline
\begin{minipage}[t]{0.08\columnwidth}\raggedright\strut
Allocation\strut
\end{minipage} & \begin{minipage}[t]{0.05\columnwidth}\raggedright\strut
HI\strut
\end{minipage} & \begin{minipage}[t]{0.19\columnwidth}\raggedright\strut
Potential harvest index\strut
\end{minipage} & \begin{minipage}[t]{0.06\columnwidth}\raggedright\strut
-\strut
\end{minipage} & \begin{minipage}[t]{0.06\columnwidth}\centering\strut
Yes\strut
\end{minipage} & \begin{minipage}[t]{0.08\columnwidth}\centering\strut
No\strut
\end{minipage} & \begin{minipage}[t]{0.05\columnwidth}\centering\strut
No\strut
\end{minipage} & \begin{minipage}[t]{0.04\columnwidth}\centering\strut
No\strut
\end{minipage} & \begin{minipage}[t]{0.04\columnwidth}\raggedleft\strut
0.4\strut
\end{minipage} & \begin{minipage}[t]{0.04\columnwidth}\raggedleft\strut
0.25\strut
\end{minipage} & \begin{minipage}[t]{0.04\columnwidth}\raggedleft\strut
0.48\strut
\end{minipage}\tabularnewline
\begin{minipage}[t]{0.08\columnwidth}\raggedright\strut
Allocation\strut
\end{minipage} & \begin{minipage}[t]{0.05\columnwidth}\raggedright\strut
OC\strut
\end{minipage} & \begin{minipage}[t]{0.19\columnwidth}\raggedright\strut
Potential seed oil content\strut
\end{minipage} & \begin{minipage}[t]{0.06\columnwidth}\raggedright\strut
\(\%\)\strut
\end{minipage} & \begin{minipage}[t]{0.06\columnwidth}\centering\strut
Yes\strut
\end{minipage} & \begin{minipage}[t]{0.08\columnwidth}\centering\strut
No\strut
\end{minipage} & \begin{minipage}[t]{0.05\columnwidth}\centering\strut
No\strut
\end{minipage} & \begin{minipage}[t]{0.04\columnwidth}\centering\strut
Yes\strut
\end{minipage} & \begin{minipage}[t]{0.04\columnwidth}\raggedleft\strut
55\strut
\end{minipage} & \begin{minipage}[t]{0.04\columnwidth}\raggedleft\strut
48\strut
\end{minipage} & \begin{minipage}[t]{0.04\columnwidth}\raggedleft\strut
61\strut
\end{minipage}\tabularnewline
\bottomrule
\end{longtable}

\normalsize

\begin{quote}
\textbf{Table 2. Phenotypic traits used as crop model inputs and the way
to estimate them routinely.} Mean, maximum, and minimum values observed
on 89 cultivars phenotyped since 2008 are indicated.
\end{quote}

Most of these parameters are directly measured at field level in
microplots (\(\sim\) 30 m\textsuperscript{2}) or in controlled
conditions (greenhouse or outdoor platform) on isolated plants. Some
parameters are indirectly estimated from observed traits: floral
initiation and beginning of grain-filling dates are determined from
flowering date; light extinction coefficient is estimated as a function
of morphological parameters (plant height, leaf number, largest leaf
height and size). Of course, the measured traits show phenotypic
plasticity and thus their values change from site to site. We
hypothesized that phenotypic information could be used as genotypic
information (Casadebaig et al., 2011) after (1) assessing the trait
phenotypic plasticity impact on variety rankings between sites and (2)
using either the mean or maximum trait value as parameter value
(depending on the parameter meaning), if rankings were not significantly
affected (using Kendall's W test). Globally, these traits were stable
enough to be used as genotype-dependent parameters (see Casadebaig et
al., 2008 for response traits and Casadebaig (2008) p.~142-143 for
phenological and morphological traits).

The soil is simply described by water holding capacity (mm) on the soil
depth usually explored by roots and by the nitrogen mineralization rate
(kg N per normalized day at 15°C). Daily weather used for simulation is
composed of 5 common variables: maximum and minimum air temperatures (T,
°C), precipitation (P, mm), potential evapotranspiration (PET, mm),
global radiation (GR, MJ m\textsuperscript{-2}).

Crop management is described by sowing date, plant density, timing and
amount of nitrogen fertilization and irrigation. Detailed algorithm and
equations of SUNFLO can be found in Casadebaig et al. (2011) and Lecoeur
et al. (2011). The oil model was recently refined by Andrianasolo et al.
(2014). SUNFLO was first developed on a commercial modeling platform
(ModelMaker®) then it was implemented on the RECORD modeling platform
from INRA (Bergez et al., 2013). Additional documentation is also
available in the associated
\href{https://github.com/picasa/rsunflo}{rsunflo} R package (Casadebaig,
2013).

\subsubsection{Step 1: Phenotyping and model
parameterization}\label{step-1-phenotyping-and-model-parameterization}

The estimation of growth and development parameters of SUNFLO is based
on the direct measurement at field level of agronomic variables. The
parameters representing the plant response to water deficit are rather
measured in controlled conditions. The following phenotyping protocols
were implemented on the varieties evaluated in the METs.

\paragraph{Field}\label{field}

In field, dense stands (6-7 plants m\textsuperscript{-2}) were
established at conventional sowing date on microplots of 30
m\textsuperscript{2} replicated three times and well protected from
birds, weeds, and diseases. Non limiting conditions for N and water were
targeted. Two distinct types of field experiments were necessary. A
first kind of experiment was carried out on a deep soil to estimate crop
phenology and maximal leaf area development, provided that water and
nitrogen were fully available for plant until anthesis. Practically, the
experiments were conducted since 2008 on the \emph{En Crambade}
experimental station of \emph{Terres Inovia} (Montesquieu-Lauragais,
Haute-Garonne, latitude: 43.416 N, longitude: 1.629 E, altitude: 233m)
on a deep clay soil where non-limiting conditions are observed each
year. A second kind of experiment was carried out to limit vegetative
growth before anthesis (through shallow soil). Then, irrigation at
flowering allowed to maximize harvest index and oil concentration. The
experiments were conducted since 2008 at the \emph{Chambon} experimental
station of \emph{Terres Inovia} (Surgères, Charente-Maritime, latitude:
46.109 N, longitude: 0.752 W, altitude: 45 m) on a shallow, calcareous
soil (\emph{Groies}).

The phenological stages were regularly scored (emergence, early
flowering, physiological maturity). At flowering, the following
variables were measured for 5 plants per replicate: total leaf number
(TLN), leaf area (LLS) and position of the largest leaf from the bottom
(LLH), and plant height (for the estimation of the light extinction
coefficient). At physiological maturity, 10 plants were sampled per plot
for measuring the potential harvest index (HI). Achene oil concentration
was determined by Nuclear Magnetic Resonance by \emph{Terres Inovia}
national laboratory in Ardon. Potential oil content (OC) was determined
as the 9th decile of the distribution of oil concentration values
measured in METs by \emph{GEVES} and \emph{Terres Inovia} during the
pre- and post-registration process at national level.

\paragraph{Controlled conditions}\label{controlled-conditions}

Following Sinclair and Ludlow (1986) approach, we used a protocol in
controlled conditions (Lecoeur and Sinclair, 1996) to determine the
response of leaf expansion and transpiration at the plant scale after
stopping watering and leaving the soil progressively drying (dry-down
design) (Casadebaig et al., 2008). The objective was to monitor the
response of ecophysiological variables (plant transpiration, stomatal
conductance, leaf expansion) to increasing water deficit. For each
tested genotype, a logistic model was proposed to describe the plant
response (equation \ref{eq:response}).

\begin{equation}
 y = -1 + \frac{2}{1 + exp^{(a \times x)}}
 \label{eq:response}
\end{equation}

with \(y\), relative plant transpiration rate or relative leaf area
expansion rate (relatively to well irrigated control) and \(x\), soil
water deficit (indicated by the fraction of transpirable soil water,
FTSW). The fitting of coefficient \(a\) gives the genotype-dependent
parameter of plant response to soil water deficit used in the SUNFLO
model (hereafter named \(LE\) for leaf expansion and \(TR\) for
transpiration).

\subsubsection{Step 2: Model evaluation}\label{step-2-model-evaluation}

\paragraph{Data}\label{data}

We used the data available in the French post-registration MET from
\emph{Terres Inovia} to evaluate the predictive quality of SUNFLO. For
this proof of concept, we focused on 52 locations in 2009 to compare
measured and simulated oil yields with SUNFLO (Figure 3). Most of the
locations came from Poitou-Charentes (16), Centre (9), Midi-Pyrénées (8)
and Pays de Loire (7) regions, which represent about 75 \% of the
cultivated sunflower areas in France (Figure 3). Other regions covered
by the network were Aquitaine (3), Auvergne (2), Languedoc-Roussillon
(2), Rhône-Alpes (2), Provence-Alpes-Côte d'Azur (2) and Burgundy (1).
In each location, one to four variety trials were conducted,
corresponding to linoleic, oleic, early- or late-maturing panels of
varieties, for a total of 80 trials performed over the network in 2009
(summarised in Table S1). Only the locations that could be reasonably
described (nearby weather station, sufficient information on soil depth,
reliable information on crop management) were kept for the evaluation
step (80 out of 99 trials in the MET). Depending on the number of trials
and the number of varieties that were sown, from 6 up to 26 varieties
were compared on each location for a total of 35 distinct varieties
tested over the MET. In each trial, measured variables were pooled from
3-4 replicates. Globally, 568 average plots (variety \(\times\) trial)
were used for model evaluation on oil yield.

The data from the most representative weather stations were used as
daily input data. At the national level, the 2009 growing season was
characterized by rainfall shortage during grain filling period with
contrasted impacts on yield according to soil depth and climatic area.
This resulted in a climatic water deficit of 138 to 523 mm (sum of
precipitation minus evapotranspiration). National grain yield was 2.4 t
ha\textsuperscript{-1} in 2009 with variations from 1.9 to 2.9 t
ha\textsuperscript{-1} between 1989 and 2014. Soil water capacity was
estimated from European Soil Database Derived data (Hiederer, 2013) and
\emph{in situ} soil profiles and analysis when it was available.
Available soil water content was ranging from 80 to 230 mm. Soil water
content at sunflower planting was initialized at 80 \% of available soil
water content (based on soil analysis data before the sowing date).
Sowing date was ranging from March, 26 to May, 7. Extreme values of
plant densities were 4.8 and 6.5 plants m\textsuperscript{-2}.
Supplemental irrigation (\textless{} 60 mm) was applied only in 5
locations out of 52. The amount of nitrogen fertilizer applied was
ranging from 0 to 92 kg N ha\textsuperscript{-1}. As residual N was
generally not measured in most of the situations, a default value was
fixed at 60 kg N ha\textsuperscript{-1}. This corresponds to the average
value of N mineral at the end of winter simulated by EPICLES model in
more than 200 farmer's fields from South-West France during two years
(2007-2008) (Champolivier et al., 2011).

\includegraphics[width=0.7\textwidth]{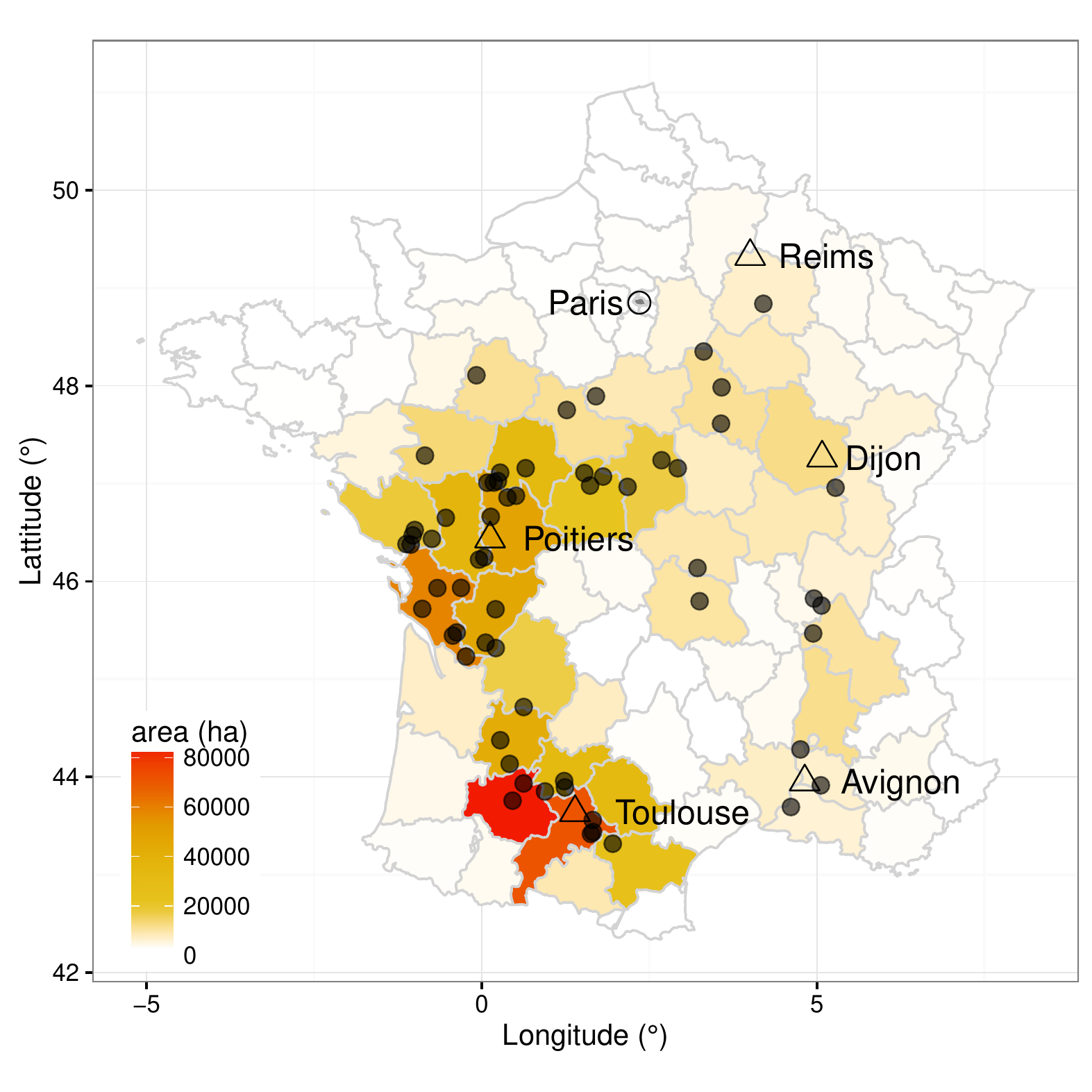}

\begin{quote}
\textbf{Figure 3. Distribution of the 52 experimental platforms from the
post-registration network from CETIOM used for SUNFLO evaluation in
2009.} The geographical distribution of experimental platforms (dots) is
mapped in relation with the main sunflower producing regions in France.
The five locations used in the numerical experiment are indicated with
triangles (from North to South : Reims, Dijon, Lusignan, Avignon,
Toulouse).
\end{quote}

\paragraph{Methods}\label{methods}

We provided a range of metrics allowing to get an overview of the crop
model performance in this specific usecase, i.e.~where input data and
observed variables are more prone to uncertainty than in designed
research trials. We evaluated the model prediction capacity using three
approaches. First, a visualization of residuals (observed - simulated
values) for environmental and genotypic main effects, along with the
computation of several goodness-of-fit metrics (RMSE, bias, Kendall's
\(\tau\)) (as suggested in Wallach et al., 2014). We also evaluated the
model capacity to rank each individual situations into three mutually
exclusive yield class (low, medium, high OY), based on 3-quantiles of
yield distribution. We derived model accuracy (the proportion of exact
predictions) and weighted Cohen's kappa (Cohen, 1968) from the
contingency table (confusion matrix, i.e.~counts of observed class as a
function of predicted class). Cohen's \(\kappa\) is a measure of
agreement between two qualitative variables (observation, simulation).
Finally, to focus on G\(\times\)E interactions we compared the model
error with three kind of parameterizations, expressed as (1)
\(Y = f(E)\), i.e.~averaging genetic variability; (2) \(Y = f(G)\),
i.e.~averaging environmental variability and (3) \(Y = f(G, E)\),
i.e.~actual parameterization. This last evaluation was carried both on
this study dataset (\emph{extension}) and a previous dataset where more
distinct cultivars were evaluated on less contrasted locations
(\emph{genetics}) (described in Casadebaig et al., 2011).

\subsubsection{Step 3: Model
application}\label{step-3-model-application}

We designed a numerical experiment to illustrate the potential use of
simulation to evaluate variety performance under different environments
and management options. Five climatic stations covering the sunflower
growing area were selected: Reims, Dijon, Lusignan, Avignon, Toulouse
(Figure 3). 35 years (1978-2012) of daily climate data were used for
representing climatic variability. Soil variation in each region was
summarized by two soil depths corresponding to values of available soil
water content of 100 and 200 mm. Options of crop management were
simplified: two different sowing dates (April 1 ; April 30) and a range
of plant densities observed in practice (3, 5 and 7 plants
m\textsuperscript{-2}). Each of the 35 phenotyped varieties were
simulated on 350 pedo-climatic environments (5 locations \(\times\) 2
soils \(\times\) 35 years) on which 6 crop management options were
applied. The whole exercise resulted in 2100 virtual trials and 73500
model runs.

Our operational aim was to provide recommendations for the cultivated
genetic material according to broad cultivation conditions in order to
reduce the phenotype-environment mismatch. For that, the target
population of environments was grouped in 10 broad environments,
corresponding to location \(\times\) soil conditions. Data from the
simulated multi-environment trial was summarized by ranking the
varieties according to their mean oil yield (averaged over 35 years and
6 management conditions), for each of these 10 environments.
Additionally, each environment was characterized by mean climatic water
deficit (ET:PET ratio, \%) and mean performance level (oil yield, t
ha\textsuperscript{-1}). Concerning recommendations for coupled
variety-management options, we proceeded by ranking management options
for each broad environment \(\times\) variety combinations (350 cases).
We presented these results with a subset of five contrasted varieties
from different seed companies (ES Biba, Extrasol, NK Kondi, SY Listeo,
Vellox) and two broad environments (North deep soil, South shallow
soil). Kendall's coefficient of concordance (Kendall, 1948) was used to
test the agreement in variety or management rankings among the 10
environments.

\subsection{Results}\label{results}

\subsubsection{Step 1: Phenotyping and model
parameterization}\label{step-1-phenotyping-and-model-parameterization-1}

\paragraph{Phenotypic variability among a range of commercial
varieties}\label{phenotypic-variability-among-a-range-of-commercial-varieties}

In 2009, the performance of 35 varieties (oleic and linoleic) was tested
at field level over a range of environments in France (post-registration
MET from \emph{Terres Inovia}). The variety panel included 12 control
varieties and 23 newly registered varieties. The crop model parameters
for this panel were previously obtained in 2008 and 2009, in dedicated
experiments in field (Debaeke et al., 2010) and controlled conditions
(Casadebaig et al., 2008).

We have illustrated (in Table 2 and Figure 4) the phenotypic variability
for the crop model parameters with all the varieties phenotyped so far
(89 cultivars since 2008). If differences in phenology and achene oil
concentration were expected (Figure 4, panels A and D), less information
was available on the components of plant leaf area (TLN, LLH, LLS) and
on potential harvest index (HI) on newly released varieties. Considering
traits related to plant leaf area, leaf number ranged from 22 to 37,
with different position or size for the largest leaf. The largest leaf
was positioned from nodes 14 to 21 corresponding to two contrasting
canopy morphologies referred as \emph{fir tree} (largest leaves at the
middle-bottom) or \emph{parasol pine} (largest leaves at the middle-top)
respectively (Triboi et al., 2004). This variability was modeled by a
wide range of leaf area profiles (Figure 4, panel B). The potential
harvest index ranged from 0.33 to 0.48 (Table 2). The response of
physiological processes to water deficit illustrates two contrasted
strategies observed among recent varieties: a \emph{conservative}
strategy, where the plants react to drought stress by reducing leaf
expansion and closing their stomata when FTSW is still relatively high,
and a \emph{productive} strategy, whereby the crop keeps expanding and
transpiring despite increasing drought (Sinclair and Muchow, 2001). This
difference in genotypic response may induce G \(\times\) E interactions
depending on timing and duration of drought scenario (Casadebaig and
Debaeke, 2012), e.g.~the \emph{productive} strategy may deplete soil
water too rapidly and expose the crop subsequent water deficit in
drought-prone conditions.

\includegraphics[width=0.7\textwidth]{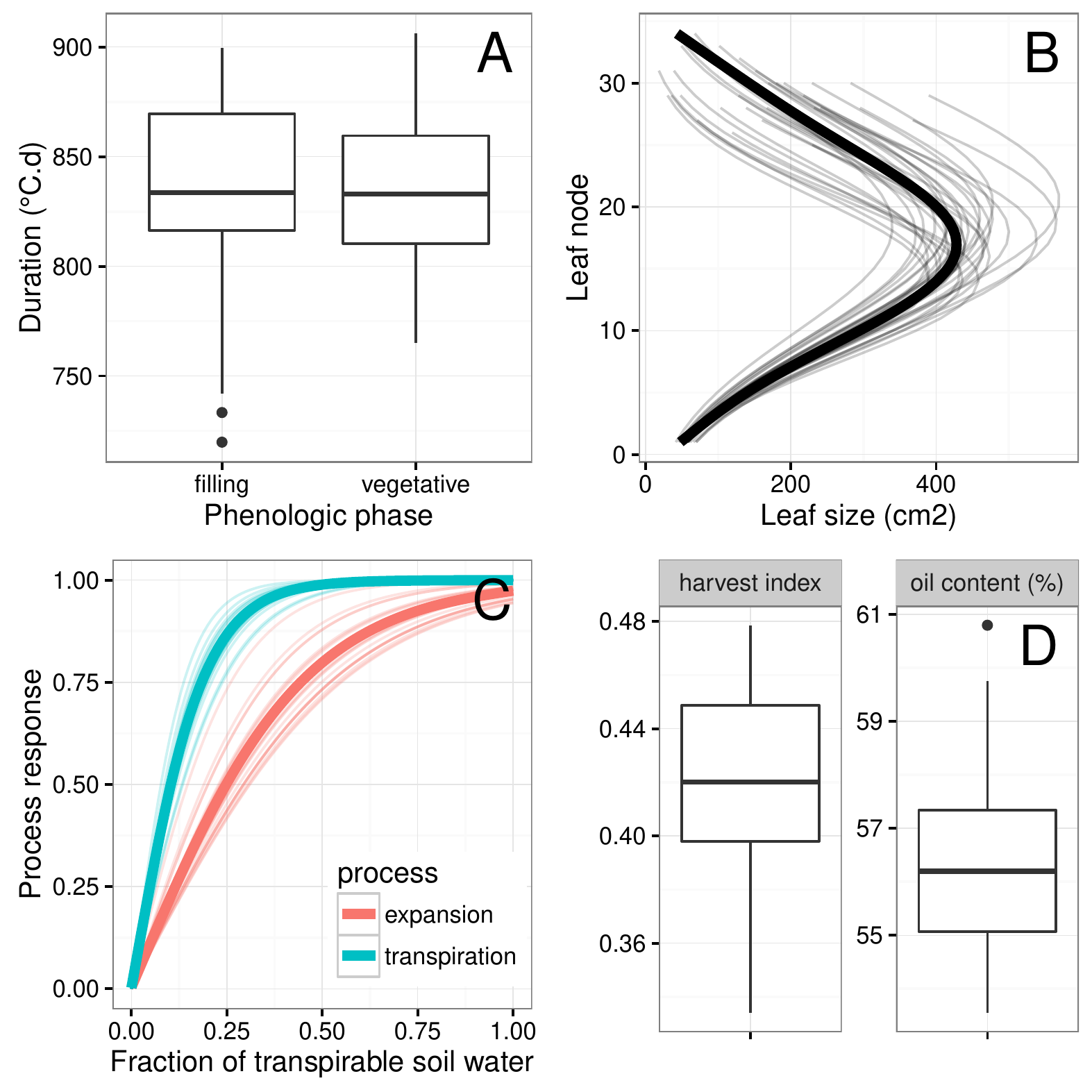}

\begin{quote}
\textbf{Figure 4. Phenotypic diversity in recent cultivated hybrids.}
Parameters are grouped by main physiological processes and we
represented either their distribution or their effect in the crop model.
Panel A represent crop development expressed in thermal time from
emergence to phenologic phase (\(TDF1\), \(TDM3\)). Panel B represent
the genotypic variability of leaf spatial distribution on stem (\(TLN\),
\(LLH\), \(LLH\)). Panel C represent the genotypic variability in plant
transpiration (blue) and expansion (red) response to water deficit.
Panel D represent potential biomass allocation to grain (\(HI\)) or
potential oil content (\(OC\))
\end{quote}

\subsubsection{Step 2: Model evaluation on post-registration
MET}\label{step-2-model-evaluation-on-post-registration-met}

The SUNFLO model was run on each of the 80 trials of the 2009 MET for
the varieties grown in each location. In this study, the evaluation of
the model performance focused on oil yield (grain yield \(\times\) oil
concentration) as it corresponds to the final commercial product and is
the most integrated variable available. The model was evaluated on its
ability to characterize abiotic stress, reproduce main environmental and
genotypic effects (Figure 6) and simulate G \(\times\) E interactions
(Figure 7).

\paragraph{Environmental
characterization}\label{environmental-characterization}

Water availability is the main limiting factor of sunflower crop in
France (Champolivier et al., 2011; Merrien, 1992; Quere, 2004). We used
the SUNFLO model to simulate water stress dynamics for each variety x
environment combination, from which we computed the actual to potential
evapotranspiration ratio (ET:PET) over the three growth period
(vegetative, flowering, grain filling) as an indicator of water stress
at the plant level. Because this indicator depends on genotype (because
of phenology, architecture and/or water response differences as
determined above), we used the mean value of ET:PET ratio for the
varieties present in the trial.

\includegraphics{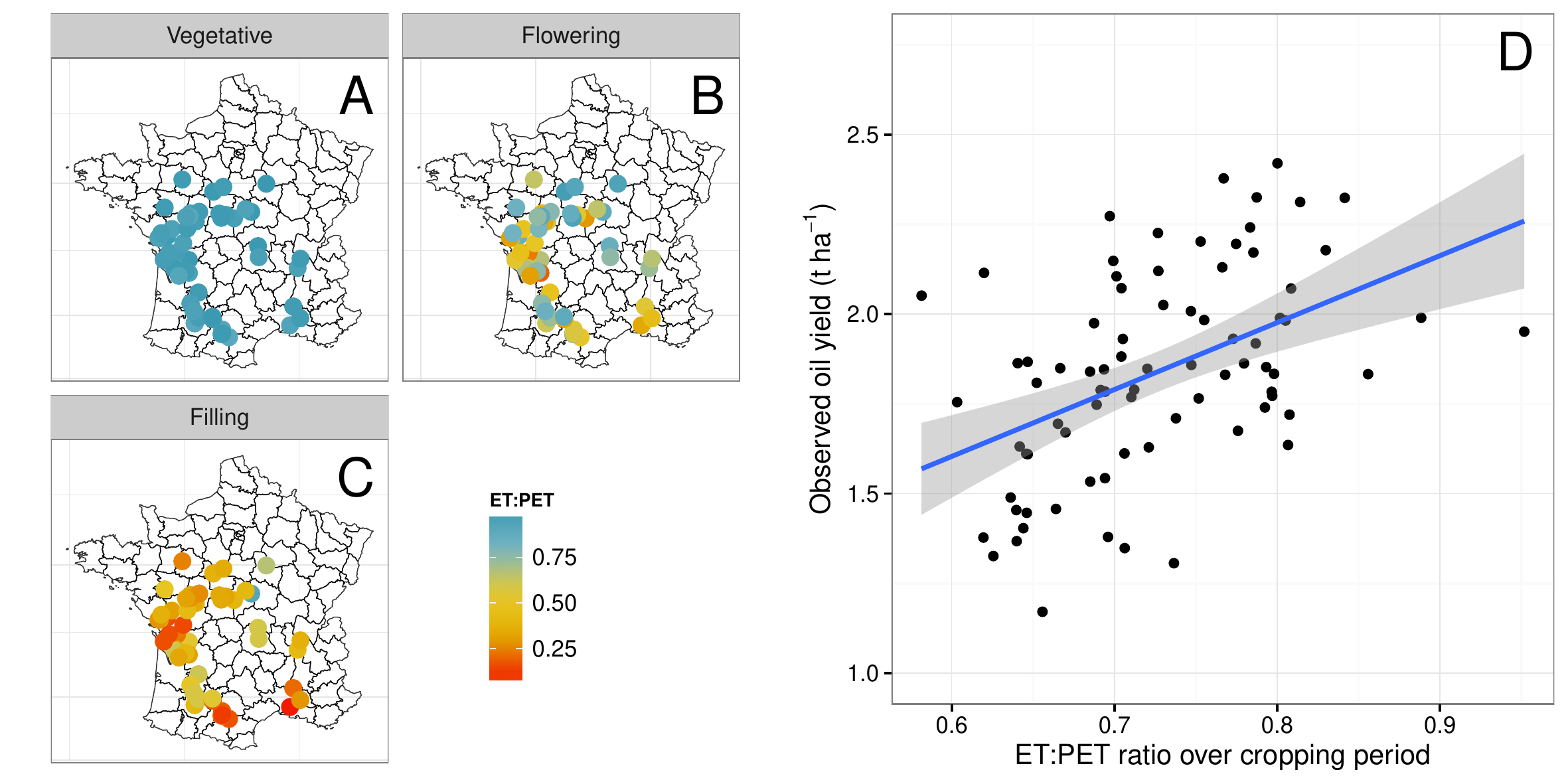}

\begin{quote}
\textbf{Figure 5. Characterization of the water stress level across the
multi-environment network.} The SUNFLO crop model was used to simulate
water stress dynamics for each genotype x environment combination and we
computed the mean ET:PET ratio over the considered growth period as an
indicator of water stress. Maps in panels A-C represent the location of
trials and the water stress level experienced by varieties during the
main crop periods: vegetative (A), flowering (B) and grain filling (C).
Panel D shows the correlation (r = 0.47, *** P \textless{} 0.001)
between observed oil yield and simulated water stress index, for each
trial.
\end{quote}

Figure 5 illustrates the distribution of water stress patterns at
national level. In 2009, as a result of differences in weather, soil and
management, ET:PET ranged from 0.58 to 0.95 when considering the whole
cropping period. Water stress progressively settled during pre-flowering
and contrasting water stress intensities were observed during flowering
and grain filling. Oil yield was significantly correlated to simulated
ET:PET (r = 0.47 ; *** P \textless{} 0.001) when considering the 80
trials. This suggests that mean oil yield (OY) per trial can be used as
an indicator for estimating model performance for ranking environments.

\paragraph{Environmental and genotypic main
effects}\label{environmental-and-genotypic-main-effects}

The ability of SUNFLO model to rank environments and sunflower varieties
is illustrated on Figure 6 where model residuals (observed minus
simulated values) are plotted against simulated values for each trial
(panel A) or varieties grown in 2009 (panel B). We used four metrics to
evaluate prediction capacity: the root-mean-square error (RMSE), the
relative RMSE (RMSE divided by observed mean), bias and Kendall's rank
correlation coefficient which measures the similarity between simulated
and observed rankings. Concerning environmental effects (panel A),
relative RMSE was 13.1 \% and Kendall's \(\tau\) was 0.48 (*** P
\textless{} 0.001). The model residuals were unbalanced, indicating a
global underestimation of oil yield (positive biais of 0.13 t
ha\textsuperscript{-1}) but no particular structure can be identified.
The simulated yield range was lower than the observed one: the simulated
yield standard deviation was 32\% lower than the observed one.
Concerning genotypic effects (panel B), relative RMSE was 11 \% and
Kendall's \(\tau\) was 0.41 (** P \textless{} 0.01). The model residuals
were unbalanced with a linear structure, indicating a systematic error,
i.e.~underestimation decreased with the level of cultivar performance
(positive biais of 0.15 t ha\textsuperscript{-1}). In both cases, the
prediction error was accurate enough to allow a significant ranking
(Kendall's \(\tau\)) of varieties or environments across the MET.

Moreover, we found a weak correlation between oil yield prediction error
per environment and distance between trial and climatic station (r =
0.19; P = 0.10) meaning that prediction error was slightly lower when
climate data were measured in the field neighbourhood.

\includegraphics{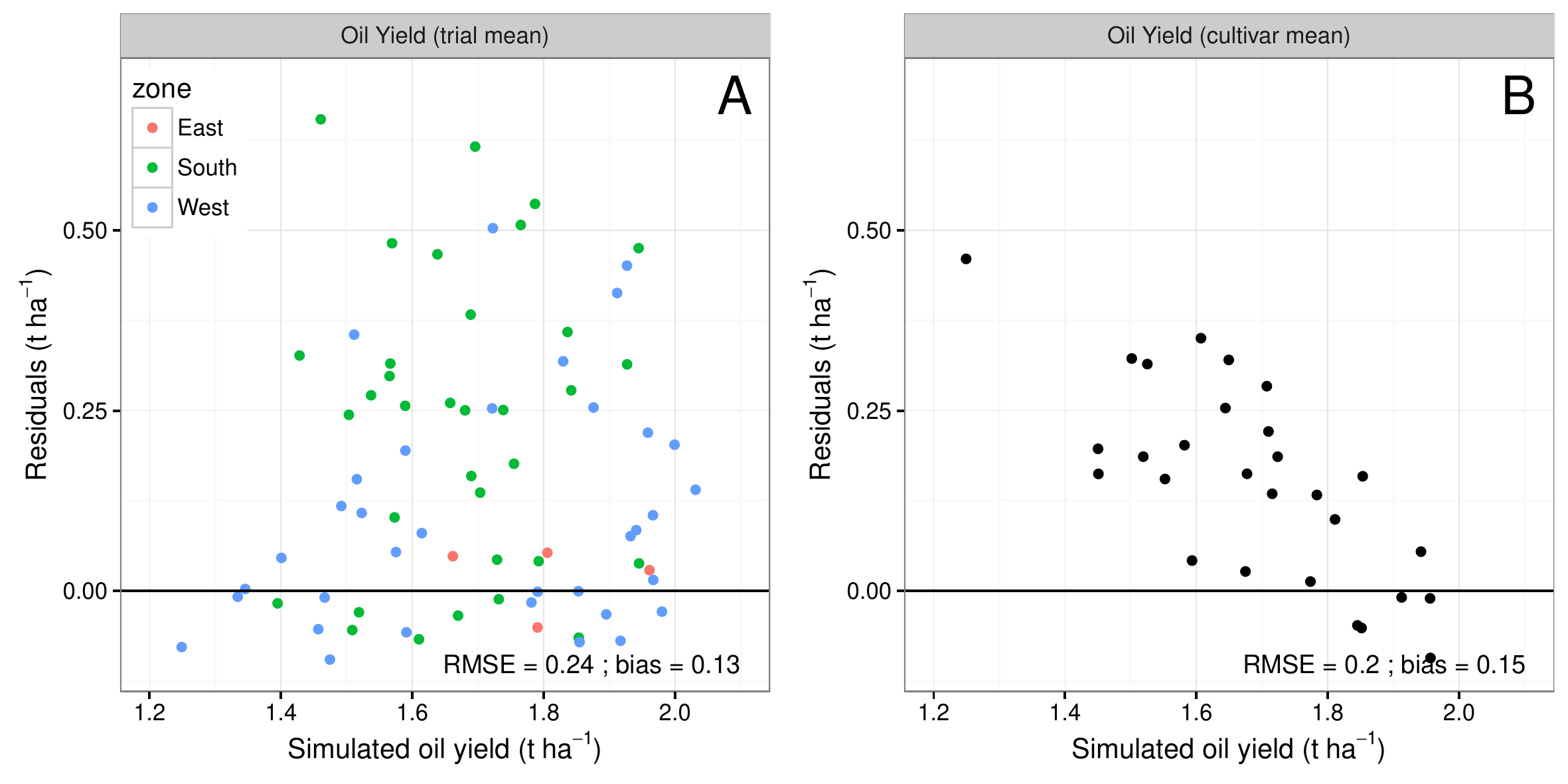}

\begin{quote}
\textbf{Figure 6. Model evaluation on an independant multi-environment
network: environmental and genotypic main effects.} In both panels,
residuals (observed values minus simulated values) are plotted against
simulated values. Panel A displays the model capacity to discriminate
between environments (average yield value per trial, n=80). Relative
RMSE was 13.1 \% ; Kendall's \(\tau\) was 0.48 (*** P \textless{}
0.001). Panel B displays the model capacity to discriminate between
genotypes (average yield value per genotype, n=28). Relative RMSE was 11
\% ; Kendall's \(\tau\) was 0.41 (** P \textless{} 0.01)
\end{quote}

\paragraph{\texorpdfstring{Genotype \(\times\) environment
interactions}{Genotype \textbackslash{}times environment interactions}}\label{genotype-times-environment-interactions}

In this evaluation approach, we used all individual situations (568
plots resulting from a subset of 35 varieties grown on 80 trials) to
evaluate the residual error (Figure 7A). Simulated oil yield ranged from
1 to 2.31 t ha\textsuperscript{-1} while observed values ranged from
0.99 to 2.6 t ha\textsuperscript{-1} with a simulated standard deviation
21 \% lower than the observed one. The model RMSE was 0.3 t
ha\textsuperscript{-1} (RRMSE = 16.4 \%) where Southern situations were
mainly responsible for under-estimation (RRMSE = 18.8 \%, biais = 0.23 t
ha\textsuperscript{-1} on trials in this subset - in green in Figure
7A). The model residuals were moderately unbalanced toward
under-estimation of oil yield without particular structure.

The confusion matrix (Figure 7B) displays the proportion of actual
G\(\times\)E combinations that were correctly predicted (diagonal) and
errors (other cells). This information was summarized with two metrics:
the model accuracy, i.e.~the proportion of exact predictions (54.4 \%)
and weighted Cohen's kappa (0.5, *** P \textless{} 0.001) (Cohen, 1968).
According to Landis and Koch (1977), a kappa value \textgreater{} 0.40
denotes a moderate agreement between two raters.

We also focused on the prediction capacity of the G\(\times\)E
interactions component. Figure S1 indicated that, in the studied MET
(\emph{extension network}), the global prediction error (RMSE) was
similar whether we used a genotype-dependent parameterization or not;
although the bias was reduced in the first case. However, it was not the
case when evaluating the model with another dataset (Casadebaig et al.,
2011), where the genotype-dependent parameterization led to more
accurate predictions (Figure S1, \emph{genetics} network).

From these agreement methods, we can conclude that SUNFLO succeeded in
representing roughly variety ranking in sunflower. Moreover, Figure 7B
also indicated that the model was able to separate the less productive
varieties from the best ones: the proportion of exact predictions was
higher in low (C) or high (A) yield class than medium class (B).

\includegraphics{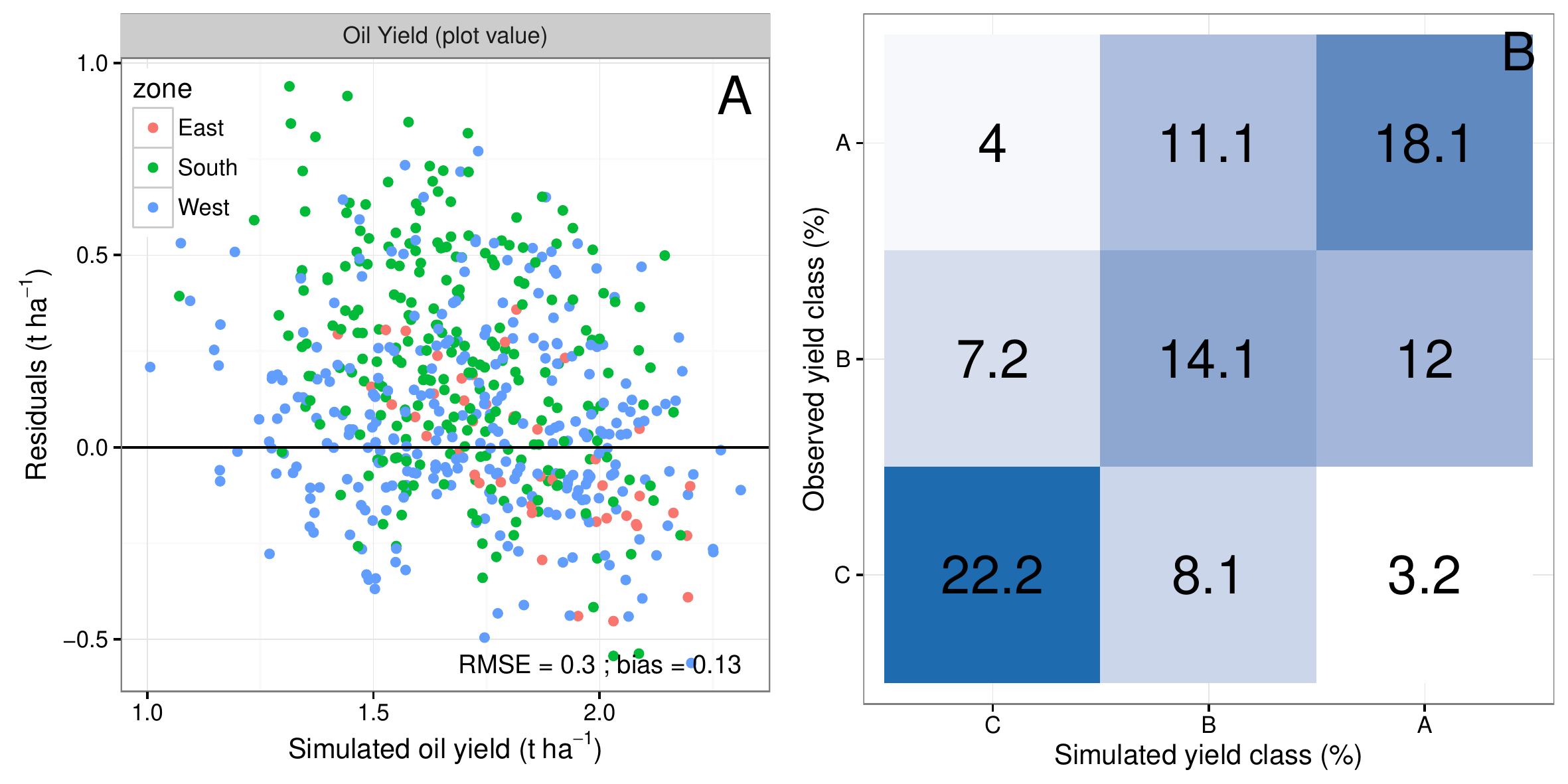}

\begin{quote}
\textbf{Figure 7. Model evaluation on an independant multi-environment
network: global quantitative and qualitative evaluation.} Panel A
displays the model prediction capacity on the complete MET (n=568
genotype \(\times\) environment combination), with colors indicating
large climatic zone in the French sunflower growing area. Relative RMSE
was 16.4 \%. Panel B displays the confusion matrix of the model capacity
to qualitatively predict crop oil yield on the same dataset. Observed
and simulated values were divided into 3 performance class from A (top
third) to C (low third). The confusion matrix represents the proportion
of actual G\(\times\)E combinations that were correctly predicted
(diagonal) and errors (other cells). Model accuracy was 54.4 \% (exact
predictions) and weighted Cohen's kappa was 0.49 (*** P \textless{}
0.001).
\end{quote}

\includegraphics[width=0.5\textwidth]{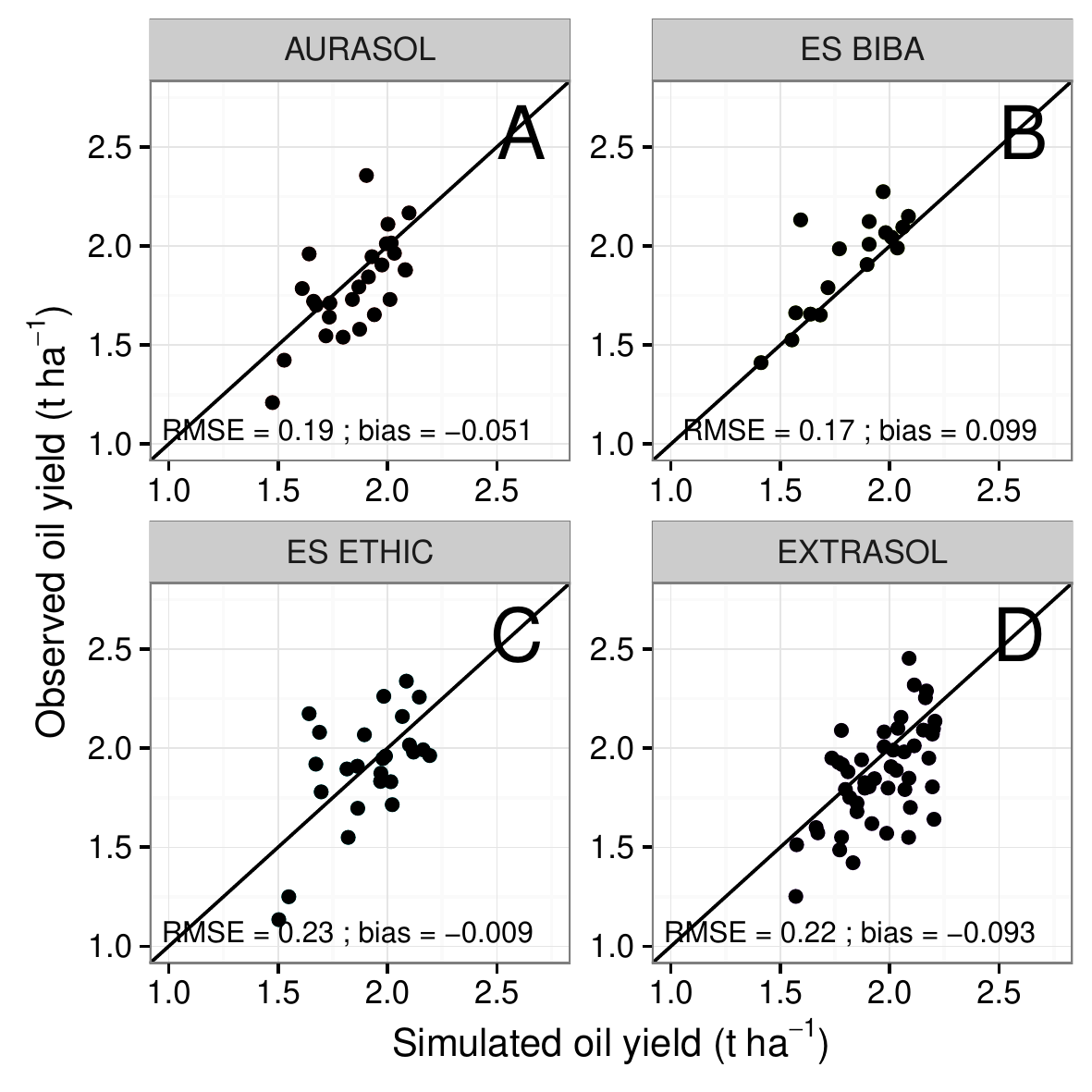}

\begin{quote}
\textbf{Figure 8. Model evaluation on an independant multi-environment
network: subset for control varieties.} Panels A-D display the model
prediction capacity for a subset of four control varieties well
represented across the MET, where each point is an individual plot. On
this subset, the relative RMSE was ranging from 9 to 12 \% ; Kendall's
\(\tau\) was ranging from 0.32 (* P \textless{} 0.05) to 0.59 (*** P
\textless{} 0.001).
\end{quote}

Finally, we focused evaluation on a subset of 4 control varieties
(\emph{Aurasol, ES BIBA, ES ETHIC, EXTRASOL}) which were more frequently
phenotyped (\(n \in [4-10]\) trials, see section \emph{Phenotyping and
model parameterization}) and more represented (\(n \in [17, 48]\))
across the MET (Figure 8, A-D) than other tested varieties. On this
subset, the prediction error was lower than in the global dataset.

\subsubsection{Step 3: Model
application}\label{step-3-model-application-1}

For this study, we proposed a simple description of the target
population of environments for sunflower in France. We used the SUNFLO
crop model to simulate a numeric multi-environment trial network for all
of the 35 phenotyped varieties, i.e.~2100 virtual trials widening the
original evaluation network of 80 trials.

\paragraph{Providing recommendation for variety
choice}\label{providing-recommendation-for-variety-choice}

Current recommendations for variety choice are mainly based on sowing
date adjustment to variety earliness, i.e.~an adaptation to the sum of
temperature available in the growing location. Our results confirmed
this trend: considering the two overall best performing varieties
(legend in Figure 9), the late-maturing one performed better in colder
environments, even if earliness was in the middle range of the
considered genotypic diversity. We also observed that the variety
rankings for the considered broad cultivation conditions were different
(Figure 9). This result illustrated that the model successfully
simulated G \(\times\) E interactions (variety ranking is different from
site to site) and could support the recommendation of varieties
performing better on specific growing condition (variety in the top row
are different). We formally tested this hypothesis by comparing the
agreement between variety rankings with Kendall's coefficient of
concordance (Kendall, 1948), which indicated a low agreement among
cultivation conditions (0.25, p = 0.038), i.e.~supported specific
recommendations. However, the quantitative differences between ranks
were more important in either dry or humid conditions than in mild
climates, where a sub-optimal variety choice had less consequences on
crop productivity (Figure 9, white figures in cells).

\includegraphics{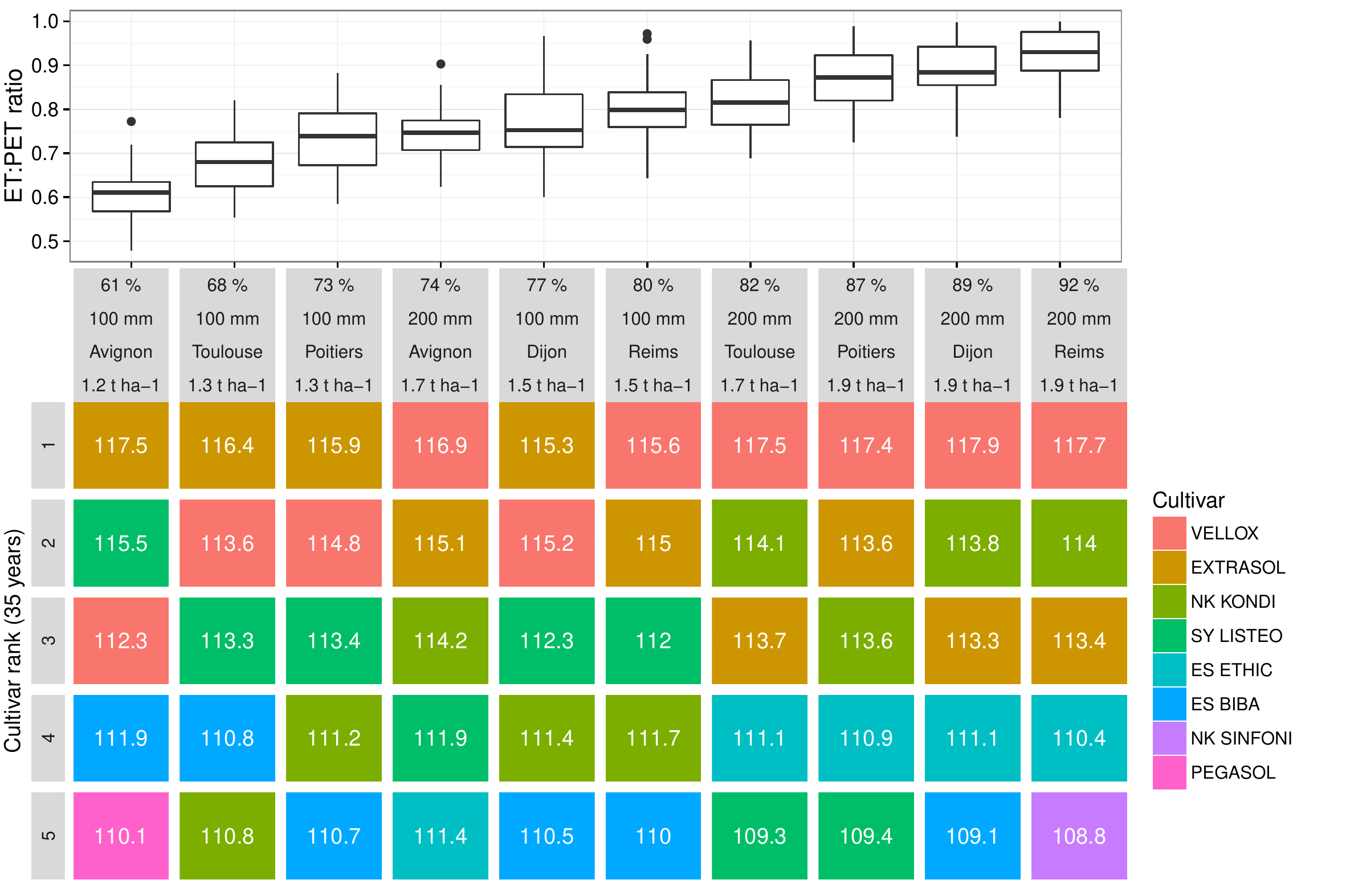}

\begin{quote}
\textbf{Figure 9. Using crop modeling to improve variety recommendations
in broad cultivation conditions} 35 varieties were ranked according to
mean crop oil yield (over 35 years), for each of the 10 site \(\times\)
soil conditions. The overall variety ranking in the TPE is indicated in
the legend. Those 10 broad cultivation conditions were sorted by
decreasing water deficit (mean ET:PET ratio) from left to right (top
boxplot). The figure presents the top five cultivars (y-axis, decreasing
performance from rank one to five) for each broad cultivation condition
(x-axis, soil water capacity, site and mean performance in column
headers). Quantitative differences between ranks are indicated in the in
cells (white figures) as variety performance relatively to mean
performance in columns (\%). Kendall's W coefficient of concordance
(Kendall, 1948) is 0.25 (* P \textless{} 0.05) indicating a low
agreement in cultivar ranking among cultivation conditions, i.e the
possibility for location-specific recommendations.
\end{quote}

\paragraph{Providing recommendations for coupled variety-management
choices.}\label{providing-recommendations-for-coupled-variety-management-choices.}

Globally, when considering all possible varieties \(\times\) broad
environments (350 cases), the management had a weak impact on crop
performance (white figures in Figure 10) and all management options were
found to be similar (W=0.15, *** P \textless{} 0.001). In this case, a
global recommendation would be to prioritize early sowings with a
planting density around 7 plants m\textsuperscript{-2}, which happens to
be similar to advices issued from agricultural extension services
(Terres Inovia, 2016).

However, when focusing on more constrasted conditions, i.e.~a subset of
five varieties from different seed companies (cultivars \emph{VELLOX,
EXTRASOL, NK KONDI, SY LISTEO, ES BIBA}) and two broad environments
(North deep soil, South shallow soil) simulation brought evidence for
linking management options to the variety choice. In figure 10, we
illustrated that management options were significantly dependent on
variety choice and cultivation conditions (W=0.29, p = 0.011), with
interactions between variety and planting density in locations more
exposed to water deficit (Toulouse, 100 mm AWC) and interactions between
varieties and sowing date in more favorable locations.

\includegraphics{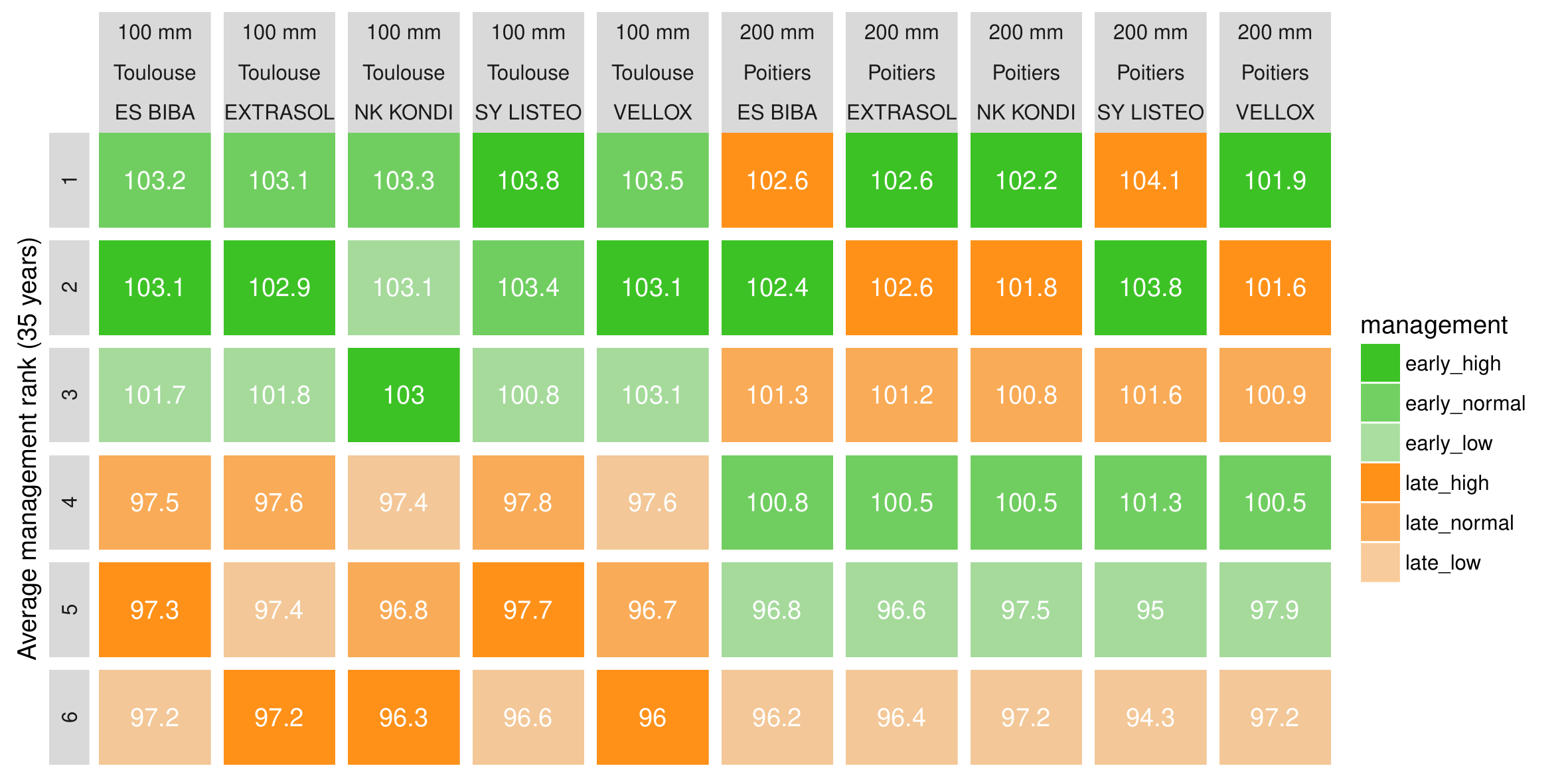}

\begin{quote}
\textbf{Figure 10. Using crop modeling to adapt crop management to
variety choice.} Six management options were ranked according to mean
crop oil yield (over 35 years), for each of the 10 site \(\times\) soil
\(\times\) variety conditions. The figure is based on a subset of five
contrasted varieties from different seed companies and two broad
environments (North deep soil - Poitier 200 mm SWC, South shallow soil -
Toulouse 100 mm SWC). Quantitative differences between ranks are
indicated in the in cells (white figures) as management performance
relatively to mean performance in columns (\%). Hue indicates early
(green) or late (orange) crop sowings, while saturation indicates the
sowing density.
\end{quote}

\subsection{Discussion}\label{discussion}

Predicting G \(\times\) E interactions with a simulation model
calibrated with measurable genotype-dependent parameters is an important
specificity of our approach. In this method, the uncertainty in
parameters values is driven solely by data rather than by both model and
data. In many other crop models, genotype-dependent parameterization
results from model fitting on intermediary or final variables (Jeuffroy
et al., 2014). Moreover, as varieties are phenotyped the same year as
their release, it allows model-based studies to keep pace with genetic
progress; a concern raised in a recent modeling impact review (Robertson
et al., 2015). We also aim to further integrate this direct
parameterization approach with the official variety evaluation chain
thereby improving transparency in model parameters.

In return, simulation could increase the efficiency of variety
evaluation because (1) phenotyping and modeling steps are adapted to the
number of varieties and trials of registration process and (2) of the
possibility to conduct numerical experiments in order to test
un-encountered situations and widen the climatic variability when
evaluating varieties.

\subsubsection{The model accuracy was sufficient to characterize
environments and rank varieties in a national evaluation
network.}\label{the-model-accuracy-was-sufficient-to-characterize-environments-and-rank-varieties-in-a-national-evaluation-network.}

The SUNFLO crop model simulated oil yield of different sunflower hybrids
in various cropping environments with a reasonable accuracy (54.4 \%)
once (1) considered varieties were phenotyped to provide
genotype-dependent parameters and (2) soil, management and climatic
conditions were correctly characterized. This performance is similar to
most of the published crop models often more complex than SUNFLO (Martre
et al., 2015; Rosenzweig et al., 2013). However, if the model
successfully ranked cultivated varieties displaying a large genetic
variability, this study pointed out a lack of of accuracy to
discriminate those closer in terms of productivity (Figure S1 and
Casadebaig et al. (2011)). The model prediction error came from the
uncertainty in the input dataset (e.g.~pedo-climatic data) but also from
the model structure, i.e.~how the physiological processes and the
limiting factors are represented or not.

The uncertainty in the characterization of trial environments impacted
prediction quality. For instance, when using the distance between trial
and climatic data measurement location as a proxy for data quality, we
showed an impact on prediction error. Representative daily climatic data
(especially precipitation) are therefore absolutely necessary. We also
made strong assumptions on soil nitrogen availability. Except for N
fertilization which was described in crop management, we considered
constant values for residual N (60 kg N ha\textsuperscript{-1}) and
potential N mineralization (0.7 kg N normalized day-1) in the absence of
information for each location. In field surveys, initial N at sunflower
sowing could range from 30 to 130 kg ha\textsuperscript{-1} N depending
of the cropping system and potential mineralization from 0.4 to 1.8 kg N
normalized day-1 (Valé et al., 2007). The use of a soil map (Hiederer,
2013) jointly with data from local soil analysis allowed to reduce
uncertainty on soil water capacity estimation. However, soil water
capacity may have been under-estimated in shallow soils with cracked
subsoil, leading to underestimation of crop performance (indicated by a
negative bias in model evaluation). As model prediction quality is
affected by inaccurate determination of water and nitrogen availability
but also initial amounts, we suggest to measure LAI on control varieties
and estimate influent input data with numerical optimisation (Debaeke et
al., 2012). Additionally, observed data are also uncertain. Experimental
design based on microplots (30 m\textsuperscript{2}) tends to
overestimate grain yield compared to farmers field conditions. A small
number of experimental location (2/54) where grain yield was over 4.5 t
ha\textsuperscript{-1} were not considered in evaluation because the
crop model was not designed for such outliers in performance level.

The modeling options, i.e.~the processes included or not in the SUNFLO
crop model also impacted prediction quality. This matter was previously
discussed in details during the model development (Casadebaig et al.,
2011) and the key points were related to the modeling of abiotic
stresses interaction and not considering biotic factors. Abiotic
stresses were represented with scalars and their interaction was modeled
with a multiplicative approach which tends to overestimate stress
effects, leading to yield underestimation in environments most exposed
to abiotic stresses (e.g.~Southern environments in Figure 7A). Biotic
factors were not considered in the model because the usual crop
management aims to limit pests damage, particularly in variety
evaluation networks. However, the occasional yield overestimation by the
model was probably due to uncontrolled limiting factors, where predation
(mainly by birds) during crop emergence and harvest may cause severe
yield losses. As with most of the crop models, there is no efficient
solution to predict yield losses caused by numerous biotic factors
(e.g.~for fungal diseases phoma, sclerotinia, phomopsis, verticillium).
Attempts to couple detailed simulation model for plant growth and pest
infection, injury and damage (Boote et al., 1983; Rouse, 1988; Savary et
al., 2000; Willocquet et al., 2008) did not yet led to tools easily
usable outside of specific pathosystems. Statistical or multi-criteria
aggregative approaches (such as Aubertot and Robin, 2013) might be used
in conjunction with crop models to represent the inherent complexity of
agroecosystems.

Overall, the SUNFLO model was able to simulate trials from a MET network
using accessible inputs (open-sourced soil map and climate data) and
without a numerical calibration step targeting genotype-dependent
parameters. On this dataset, the model performance was comparable to
other crop models calibrated and evaluated on a small number of good
experimental data (García-López et al., 2014; e.g.~for sunflower Leite
et al., 2014).

\subsubsection{Plant phenotyping to keep pace with genetic progress and
reduce uncertainty model
inputs.}\label{plant-phenotyping-to-keep-pace-with-genetic-progress-and-reduce-uncertainty-model-inputs.}

The model parameters chosen to represent varieties focused on key
physiological processes (phenology, light interception, water response)
but their number was kept compatible with manual field phenotyping
protocols. These protocols are now routinely used by \emph{Terres
Inovia} extension service since 2008 to parameterize new varieties. We
found that two environments generating contrasted growth conditions (one
for potential canopy development, one for potential biomass allocation)
are necessary and sufficient to determine the parameters of potential
growth. This experimental design could be improved by including a small
set of precisely phenotyped control varieties in each experiment to
control the environmental effect (years and locations).

However, the recent development of high-throughput plant phenotyping
could question this approach, potentially reducing uncertainty in the
measurement of genotype-dependent parameters and allowing to consider
new processes or parameters. Initially, we suggested a complete
determination of SUNFLO parameters on isolated plants grown in
greenhouse (Lecoeur et al., 2011). Although the correlation with field
plants is acceptable, considering only phenotyping platforms operating
in controlled conditions would failed to represent plant functioning in
late grain filling stages and in dense stands. Consequently, an open-air
pot platform is currently in development aiming to automatize daily
measurements for water deficit, leaf area and transpiration (Blanchet et
al., 2016). Such platform will target mainly the phenotyping of response
traits (\emph{LE}, \emph{TR}) but architectural trait will also be
evaluated. Some phenotyping methods might also be directly usable on MET
used for variety evaluation, such as using unmanned aerial vehicles and
image analysis to estimate phenological stages or multispectral camera
for canopy architecture (Baret and Buis, 2008; Verger et al., 2014).

Generally, we suggest that model calibration targeting
genotype-dependent parameters should be avoided both for conceptual and
technical reasons. From a conceptual point of view, such parameters (in
fact, phenotypic traits), are supposed to represent genotype and thus
their value should not vary among environments. Secondly, it is possible
to directly measure such parameters by limiting their number and
designing phenotyping methods during model development.

\subsubsection{Toward decision support systems based on modeling and
simulation}\label{toward-decision-support-systems-based-on-modeling-and-simulation}

The simulation of phenotypic plasticity enables a set of applications
for breeders and advisers where variety evaluation is a central process
(Jeuffroy et al., 2014). We briefly illustrated two of these
applications: (1) the characterization of water stress constraints
perceived at the plant level, in each trial environment (Figure 5) and
(2) the determination of agronomic suitability of registered varieties,
including variety-dependent management recommendations (Figures 9-10).
In the later numerical experiment, we designed a simple scenario, using
only 10 site x soil conditions to describe the actual sunflower growing
area. Nevertheless, we illustrated that (1) distinct varieties could be
recommended according to local pedo-climatic context and that (2) this
recommendation could led to crop performance increase under climatic
uncertainty (35 years of historical data). Our case study also
illustrated that even if there was evidence for a variety-dependent crop
management (sowing date and plant density), its impact on crop
performance was modest (+ 3\% compared to average management).

Simulation also enables new applications for variety evaluation. Because
recurrent field trials (Figure 1) are dependent of climate variability
(and in longer term, climate change), varieties released in different
years were not evaluated in the same conditions. By designing a virtual
MET to compare those varieties under the same climatic variability, it
is possible to compare them and to assess the rate genetic progress
independently of climate change. Before considering such applications,
we need to develop a framework linking crop simulation with genotypes,
soil and climate datasets so that involved stakeholders can design
numerical experiments adapted to their specifications. Such operational
decision tools should focus on parsimonious models and deliver
integrated outputs easy to interpret.

\section{Conclusions}\label{conclusions}

Our aim was to present a model-based approach to assist the variety
evaluation process. Although this approach was implemented on sunflower
crop, it is not limited to a specific crop if the simulation model is
adapted. We showed that few phenotypic traits used as genotype-dependent
parameters are sufficient to predict phenotypic plasticity observed for
recent hybrids tested in evaluation networks. Nevertheless, we found
that the model performed better for assessing abiotic stress (G, E
component) than for ranking cultivars (G\(\times\)E interactions
component), especially in an agricultural extension network with
inherent uncertainty in input data. Linking official variety evaluation
multi-environment trials and crop modeling allowed to amplify the
environmental and agronomic conditions in which the varieties are
routinely tested. We suggest that this approach could find operational
outcomes to recommend varieties according to environment types. Such
spatial management of genetic resources could potentially improve crop
performance by reducing the genotype-phenotype mismatch in farming
environments.

\section{Acknowledgements}\label{acknowledgements}

The authors are grateful to the students (Claire Barbet-Massin, Ewen
Gery, Bertrand Haquin) and staff from \emph{INRA}, \emph{ENSAT} (Michel
Labarrère, Pierre Maury, Colette Quinquiry) and \emph{Terres Inovia}
(Frédéric Bardy, Philippe Christante, André Estragnat, Pascal Fauvin,
Jean-Pierre Palleau, Frédéric Salvi) that helped to constitute the
phenotypic database, helped in modeling and simulation (RECORD team from
INRA, Eric Casellas, Gauthier Quesnel, Helène Raynal, Ronan Trépos) and
provided climatic datasets (AgroClim team from INRA). Grants were
provided by the French Ministry of Agriculture (AAP CTPS 2007 \& 2010),
the French Ministry of Research (ANR SUNRISE ANR-11-BTBR-0005), and the
PROMOSOL association (\emph{Productivité du Tournesol 3}).

\section{Supplementary material}\label{supplementary-material}

\footnotesize

\begin{longtable}[]{@{}rlrrrrrllrr@{}}
\toprule
\# & site & lat & lon & SCWD & AWC & density & sowing & harvest &
irrigation & fertilization\tabularnewline
\midrule
\endhead
1 & Saulzet & 46.14 & 3.22 & -138 & 121 & 5.6 & Mar-27 & Sep-07 & 0 &
85\tabularnewline
2 & Castelnaudary & 43.32 & 1.95 & -473 & 157 & 5 & May-04 & Sep-07 & 0
& 40\tabularnewline
3 & Castelnaudary & 43.32 & 1.95 & -478 & 157 & 6.5 & May-07 & Sep-07 &
0 & 60\tabularnewline
4 & Champniers & 45.71 & 0.21 & -281 & 148 & 5.6 & Apr-20 & Sep-07 & 60
& 50\tabularnewline
5 & Tusson & 45.32 & 0.21 & -283 & 86 & 6 & Apr-01 & Aug-27 & 0 &
50\tabularnewline
6 & Gibourne & 45.93 & -0.31 & -332 & 130 & 5.6 & Apr-07 & Sep-02 & 0 &
80\tabularnewline
7 & Reaux & 45.48 & -0.37 & -280 & 121 & 5.6 & Apr-25 & Sep-10 & 0 &
60\tabularnewline
8 & Sablonceaux & 45.72 & -0.89 & -282 & 98 & 6.5 & Apr-14 & Sep-08 & 0
& 60\tabularnewline
9 & Jonzac & 45.45 & -0.43 & -229 & 185 & 5.6 & Apr-06 & Sep-08 & 0 &
55\tabularnewline
10 & Civray & 46.97 & 2.17 & -247 & 73 & 5.2 & Apr-06 & Sep-09 & 0 &
46\tabularnewline
11 & Montigny & 47.24 & 2.68 & -238 & 73 & 6.5 & Apr-14 & Sep-11 & 0 &
60\tabularnewline
12 & Pecdorat & 44.71 & 0.62 & -218 & 140 & 5.6 & Mar-31 & Aug-27 & 0 &
60\tabularnewline
13 & Fourques & 43.69 & 4.61 & -523 & 158 & 4.8 & Apr-25 & Sep-07 & 0 &
42\tabularnewline
14 & Montesquieu Lauragais & 43.42 & 1.63 & -348 & 217 & 5.4 & May-12 &
Sep-03 & 0 & 80\tabularnewline
15 & Montesquieu Lauragais & 43.42 & 1.63 & -357 & 217 & 6.5 & May-05 &
Sep-03 & 0 & 60\tabularnewline
16 & Lanta & 43.56 & 1.65 & -453 & 217 & 6.5 & May-05 & Sep-17 & 0 &
60\tabularnewline
17 & Villenouvelle & 43.44 & 1.66 & -370 & 169 & 6.5 & May-05 & Aug-29 &
0 & 60\tabularnewline
18 & Jegun & 43.76 & 0.46 & -339 & 121 & 6.5 & May-05 & Sep-02 & 0 &
60\tabularnewline
19 & Jegun & 43.76 & 0.46 & -339 & 121 & 6.5 & May-05 & Sep-02 & 0 &
60\tabularnewline
20 & Vatan & 47.07 & 1.81 & -257 & 101 & 6.2 & Apr-17 & Sep-14 & 0 &
45\tabularnewline
21 & Ligré & 47.11 & 0.27 & -312 & 82 & 5.6 & Apr-02 & Sep-04 & 0 &
0\tabularnewline
22 & Ligré & 47.11 & 0.27 & -326 & 82 & 6.5 & Apr-14 & Sep-04 & 0 &
60\tabularnewline
23 & Ste Catherine De Fierbois & 47.16 & 0.65 & -347 & 158 & 6.5 &
Apr-14 & Sep-04 & 0 & 60\tabularnewline
24 & Saint Sorlin De Vienne & 45.47 & 4.94 & -415 & 189 & 5.9 & Apr-21 &
Sep-11 & 0 & 50\tabularnewline
25 & Rhodon & 47.75 & 1.27 & -265 & 158 & 6.5 & Apr-15 & Sep-11 & 0 &
60\tabularnewline
26 & Huisseau & 47.89 & 1.7 & -265 & 175 & 6.5 & Apr-06 & Sep-10 & 0 &
60\tabularnewline
27 & Villeton & 44.38 & 0.28 & -323 & 189 & 6.5 & Apr-24 & Sep-09 & 0 &
60\tabularnewline
28 & Villeton & 44.38 & 0.28 & -323 & 189 & 6.5 & Apr-24 & Sep-09 & 0 &
60\tabularnewline
29 & Calignac & 44.13 & 0.42 & -342 & 217 & 6.5 & Apr-24 & Sep-14 & 0 &
60\tabularnewline
30 & Calignac & 44.13 & 0.42 & -342 & 217 & 6.5 & Apr-24 & Sep-14 & 0 &
60\tabularnewline
31 & Pont Du Château & 45.8 & 3.25 & -277 & 106 & 6 & May-06 & Sep-25 &
0 & 40\tabularnewline
32 & Pusignan & 45.75 & 5.07 & -433 & 189 & 5.8 & Apr-07 & Aug-27 & 0 &
45\tabularnewline
33 & Tennie & 48.11 & -0.08 & -341 & 158 & 6.5 & Apr-15 & Sep-10 & 0 &
60\tabularnewline
34 & Monbequi & 43.89 & 1.24 & -365 & 118 & 6.5 & May-05 & Sep-11 & 0 &
60\tabularnewline
35 & Montech & 43.96 & 1.23 & -370 & 178 & 6.5 & May-05 & Sep-14 & 0 &
60\tabularnewline
36 & Bollene & 44.28 & 4.75 & -393 & 189 & 5.6 & Mar-31 & Aug-27 & 0 &
60\tabularnewline
37 & Bollene & 44.28 & 4.75 & -441 & 189 & 6.5 & Apr-10 & Aug-27 & 0 &
60\tabularnewline
38 & Isle Sur La Sorgue & 43.91 & 5.06 & -475 & 209 & 5.6 & Apr-09 &
Sep-08 & 0 & 40\tabularnewline
39 & Isle Sur La Sorgue & 43.91 & 5.06 & -472 & 209 & 6.5 & Apr-10 &
Sep-08 & 0 & 60\tabularnewline
40 & St Martin De Fraigneau & 46.43 & -0.74 & -341 & 148 & 5.4 & Apr-22
& Sep-10 & 0 & 50\tabularnewline
41 & Ste Radegonde Des Noyers & 46.38 & -1.06 & -329 & 171 & 5.6 &
Apr-02 & Aug-26 & 0 & 40\tabularnewline
42 & Nalliers & 46.47 & -1.03 & -330 & 119 & 5.6 & Apr-22 & Sep-10 & 0 &
20\tabularnewline
43 & Usseau & 46.88 & 0.51 & -314 & 175 & 5.7 & Apr-10 & Sep-09 & 0 &
66\tabularnewline
44 & Sossay & 46.86 & 0.38 & -311 & 121 & 5.6 & Apr-06 & Sep-10 & 0 &
60\tabularnewline
45 & Esnon & 47.98 & 3.58 & -215 & 158 & 6 & Apr-03 & Sep-08 & 0 &
50\tabularnewline
46 & St Martial & 45.37 & 0.06 & -249 & 150 & 5.6 & Apr-21 & Sep-07 & 0
& 46\tabularnewline
47 & St Martial & 45.37 & 0.06 & -254 & 150 & 6.5 & Apr-17 & Sep-07 & 0
& 60\tabularnewline
48 & Orignolles & 45.23 & -0.24 & -274 & 78 & 5.6 & Apr-23 & Sep-16 & 0
& 72\tabularnewline
49 & Sablonceaux & 45.72 & -0.89 & -335 & 98 & 5.6 & May-05 & Sep-14 &
35 & 74\tabularnewline
50 & Les Nouillers & 45.93 & -0.66 & -341 & 130 & 5.6 & Apr-21 & Sep-08
& 0 & 69\tabularnewline
51 & St Martin Des Champs & 47.16 & 2.92 & -203 & 243 & 5.8 & Mar-26 &
Aug-28 & 0 & 50\tabularnewline
52 & Montigny & 47.24 & 2.68 & -235 & 97 & 5.6 & Apr-14 & Sep-11 & 0 &
39\tabularnewline
53 & Pecdorat & 44.71 & 0.62 & -218 & 140 & 5.6 & Mar-31 & Aug-27 & 0 &
60\tabularnewline
54 & Montesquieu Lauragais & 43.42 & 1.63 & -351 & 155 & 6.5 & May-06 &
Sep-07 & 0 & 60\tabularnewline
55 & Montesquieu Lauragais & 43.42 & 1.63 & -351 & 155 & 6.5 & May-06 &
Sep-07 & 0 & 60\tabularnewline
56 & Lanta & 43.56 & 1.65 & -452 & 155 & 5.4 & May-07 & Sep-17 & 0 &
30\tabularnewline
57 & Villenouvelle & 43.44 & 1.66 & -286 & 112 & 5.3 & Apr-06 & Aug-29 &
0 & 50\tabularnewline
58 & Jegun & 43.76 & 0.46 & -294 & 112 & 5.6 & Apr-09 & Sep-02 & 0 &
69\tabularnewline
59 & Jegun & 43.76 & 0.46 & -341 & 141 & 6.5 & May-05 & Sep-02 & 0 &
60\tabularnewline
60 & Lectoure & 43.93 & 0.62 & -311 & 224 & 5.6 & Apr-09 & Aug-31 & 0 &
61\tabularnewline
61 & Levroux & 46.98 & 1.61 & -262 & 158 & 5.6 & Apr-16 & Sep-11 & 35 &
54\tabularnewline
62 & Vicq Sur Nahon & 47.11 & 1.53 & -259 & 158 & 6 & Apr-19 & Sep-14 &
0 & 36\tabularnewline
63 & Ligré & 47.11 & 0.27 & -327 & 101 & 6.5 & Apr-14 & Sep-04 & 0 &
60\tabularnewline
64 & Ste Catherine De Fierbois & 47.16 & 0.65 & -313 & 158 & 5.2 &
Apr-21 & Sep-03 & 0 & 65\tabularnewline
65 & Rhodon & 47.75 & 1.27 & -249 & 158 & 5.6 & Apr-10 & Sep-09 & 60 &
60\tabularnewline
66 & Huisseau & 47.89 & 1.7 & -276 & 175 & 5.6 & Apr-23 & Sep-16 & 0 &
54\tabularnewline
67 & Villeton & 44.38 & 0.28 & -319 & 189 & 5.6 & Apr-23 & Sep-09 & 10 &
70\tabularnewline
68 & Villeton & 44.38 & 0.28 & -323 & 189 & 6.5 & Apr-24 & Sep-09 & 0 &
60\tabularnewline
69 & Calignac & 44.13 & 0.42 & -343 & 217 & 5.6 & Apr-24 & Sep-14 & 0 &
92\tabularnewline
70 & Ste Christine & 47.29 & -0.85 & -306 & 189 & 6 & May-05 & Sep-14 &
0 & 45\tabularnewline
71 & St Vincent La Chatre & 46.22 & -0.04 & -348 & 131 & 5.4 & Apr-27 &
Sep-14 & 0 & 72\tabularnewline
72 & Sainte-Blandine & 46.65 & -0.54 & -310 & 103 & 6 & Apr-21 & Sep-09
& 0 & 50\tabularnewline
73 & Monbequi & 43.89 & 1.24 & -368 & 130 & 5.5 & May-07 & Sep-11 & 0 &
80\tabularnewline
74 & Monbequi & 43.89 & 1.24 & -366 & 130 & 6.5 & May-05 & Sep-11 & 0 &
60\tabularnewline
75 & Montech & 43.96 & 1.23 & -321 & 240 & 6 & Apr-10 & Sep-14 & 0 &
50\tabularnewline
76 & Champagne Les Marais & 46.38 & -1.12 & -318 & 94 & 5.6 & Apr-02 &
Sep-02 & 0 & 50\tabularnewline
77 & St Etienne De Brillouet & 46.53 & -1 & -333 & 189 & 5.6 & Apr-22 &
Sep-10 & 0 & 20\tabularnewline
78 & Frozes & 46.66 & 0.13 & -307 & 189 & 5.4 & Apr-22 & Sep-10 & 0 &
50\tabularnewline
79 & Ceaux En Loudun & 47.03 & 0.24 & -385 & 161 & 5.4 & May-06 & Sep-14
& 0 & 50\tabularnewline
80 & Loudun & 47.01 & 0.08 & -360 & 166 & 5.5 & Apr-15 & Sep-10 & 0 &
60\tabularnewline
\bottomrule
\end{longtable}

\normalsize

\begin{quote}
\textbf{Table S1: Description of locations and management practices on
the multi-environment trial.} Headers indicates the locations and years
of trials, the climatic water deficit (SCWD, mm) i.e.~the sum of
precipitation minus sum of potential evapotranspiration, the plant
available water capacity (AWC, mm) i.e the amount of soil water
reserves, the plant density at sowing (plants m\textsuperscript{-2}),
the sowing and harvest dates, the amount of irrigation (mm) and nitrogen
fertilization (kg ha\textsuperscript{-1} eq. mineral nitrogen).
\end{quote}

\begin{longtable}[]{@{}llrrrrr@{}}
\toprule
zone & site & lat & lon & ST & SCWD\_m & SCWD\_sd\tabularnewline
\midrule
\endhead
East & Dijon & 47.3 & 5.1 & 2646.0 & -263.5 & 113.3\tabularnewline
East & Reims & 49.3 & 4.0 & 2454.6 & -229.5 & 78.9\tabularnewline
South & Avignon & 44.0 & 4.8 & 3178.7 & -524.1 & 120.9\tabularnewline
South & Toulouse & 43.6 & 1.4 & 2958.0 & -405.2 & 135.6\tabularnewline
West & Poitiers & 46.4 & 0.1 & 2586.6 & -294.5 & 102.4\tabularnewline
\bottomrule
\end{longtable}

\begin{quote}
\textbf{Table S2: Description of locations on the simulated
multi-environment trial} Headers indicate the locations of trials; the
sum of temperatures (ST, °C), the mean climatic water deficit (SCWD\_m,
mm) and standard deviation (SCWD\_sd, mm) i.e.~the sum of precipitation
minus sum of potential evapotranspiration.
\end{quote}

\includegraphics{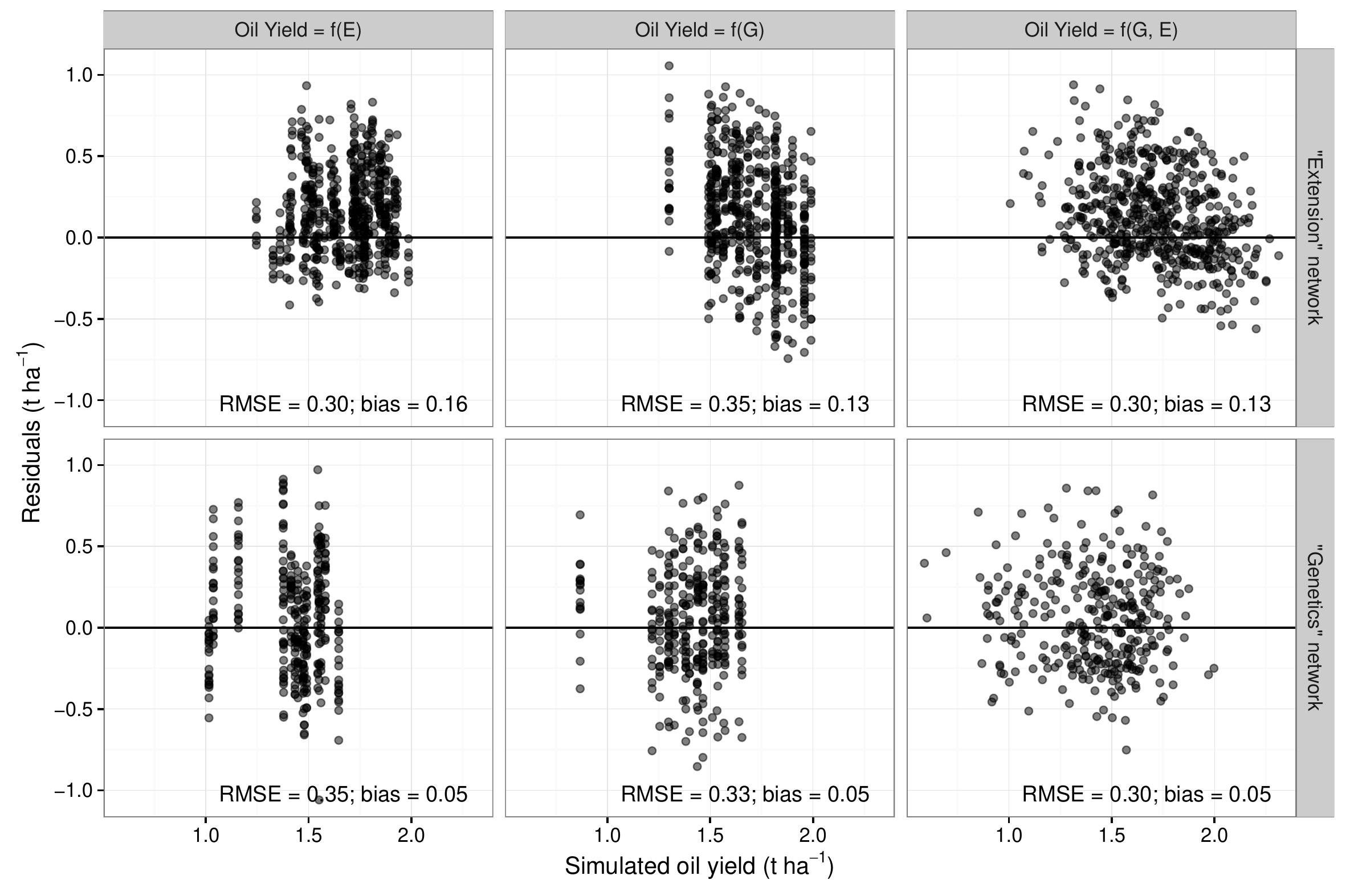}

\begin{quote}
\textbf{Figure S1: Evaluation of simulation of G\(\times\)E interactions
in two contrasted multi-environment trial networks.} We compared the
model error with three kind of parameterizations (in columns) and two
datasets (lines). Different parameterization were expressed as (1)
\(Y = f(E)\), i.e.~averaging genetic variability; (2) \(Y = f(G)\),
i.e.~averaging environmental variability and (3) \(Y = f(G, E)\),
i.e.~actual parameterization. Two dataset were used : this study
(\emph{extension}) and a MET in which more distinct cultivars were
evaluated on less contrasted locations (\emph{genetics}) (described in
Casadebaig et al., 2011).
\end{quote}

\newpage

\section*{References}\label{references}
\addcontentsline{toc}{section}{References}

\hypertarget{refs}{}
\hypertarget{ref-Andrianasolo2014}{}
Andrianasolo, F.N., Casadebaig, P., Maury, P., Maza, E., Champolivier,
L., Debaeke, P., 2014. Prediction of sunflower grain oil concentration
as a function of variety, crop management and environment by the means
of statistical models. European Journal of Agronomy 54, 84--96.
doi:\href{https://doi.org/10.1016/j.eja.2013.12.002}{10.1016/j.eja.2013.12.002}

\hypertarget{ref-Aubertot2013}{}
Aubertot, J.-N., Robin, M.-H., 2013. Injury Profile SIMulator, a
qualitative aggregative modelling framework to predict crop injury
profile as a function of cropping practices, and the abiotic and biotic
environment. i. conceptual bases. PLoS one 8, e73202.
doi:\href{https://doi.org/10.1371/journal.pone.0073202}{10.1371/journal.pone.0073202}

\hypertarget{ref-Baret2008}{}
Baret, F., Buis, S., 2008. Estimating canopy characteristics from remote
sensing observations: Review of methods and associated problems, in:
Advances in Land Remote Sensing. Springer, pp. 173--201.
doi:\href{https://doi.org/10.1007/978-1-4020-6450-0_7}{10.1007/978-1-4020-6450-0\_7}

\hypertarget{ref-Bergez2013}{}
Bergez, J., Chabrier, P., Gary, C., Jeuffroy, M., Makowski, D., Quesnel,
G., Ramat, E., Raynal, H., Rousse, N., Wallach, D., Debaeke, P., Durand,
P., Duru, M., Dury, J., Faverdin, P., Gascuel-Odoux, C., Garcia, F.,
2013. An open platform to build, evaluate and simulate integrated models
of farming and agro-ecosystems. Environmental Modelling \& Software 39,
39--49.
doi:\href{https://doi.org/10.1016/j.envsoft.2012.03.011}{10.1016/j.envsoft.2012.03.011}

\hypertarget{ref-Blanchet2016}{}
Blanchet, N., Casadebaig, P., Burger, P., Vares, D., Colombet, C.,
Boniface, M.-C., Vincourt, P., Debaeke, P., Langlade, N., 2016.
HELIAPHEN : A high-throughput phenotyping platform to characterize plant
responses to water stress from seedling stage to seed set, in: 19th
International Sunflower Conference, 29 May - 3 June, Edirne, Turkey.

\hypertarget{ref-Boote1983}{}
Boote, K., Jones, J., Mishoe, J., Berger, R., 1983. Coupling pests to
crop growth simulators to predict yield reductions. Phytopathology 73,
1581--1587.
doi:\href{https://doi.org/10.1094/phyto-73-1581}{10.1094/phyto-73-1581}

\hypertarget{ref-Brisson2003}{}
Brisson, N., Gary, C., Justes, E., Roche, R., Mary, B., Ripoche, D.,
Zimmer, D., Sierra, J., Bertuzzi, P., Burger, P., Bussière, F.,
Cabidoche, Y.M., Cellier, P., Debaeke, P., Gaudillère, J.P., Hénault,
C., Maraux, F., Seguin, B., Sinoquet, H., 2003. An overview of the crop
model STICS. European Journal of Agronomy 18, 309--332.
doi:\href{https://doi.org/10.1016/s1161-0301(02)00110-7}{10.1016/s1161-0301(02)00110-7}

\hypertarget{ref-Bustos-Korts2016}{}
Bustos-Korts, D., Malosetti, M., Chapman, S., Eeuwijk, F. van, 2016.
Modelling of genotype by environment interaction and prediction of
complex traits across multiple environments as a synthesis of crop
growth modelling, genetics and statistics, in: Crop Systems Biology.
Springer, pp. 55--82.

\hypertarget{ref-Casadebaig2013}{}
Casadebaig, P., 2013. rsunflo, a R package for phenotyping, simulating
and modelling with the SUNFLO crop model. INRA.

\hypertarget{ref-Casadebaig2008b}{}
Casadebaig, P., 2008. Analyse et modélisation de l'interaction Génotype
- Environnement - Conduite de culture: Application au tournesol
(Helianthus annuus L.) (PhD Thesis). Toulouse University.
doi:\href{https://doi.org/10.6084/m9.figshare.787695}{10.6084/m9.figshare.787695}

\hypertarget{ref-Casadebaig2012b}{}
Casadebaig, P., Debaeke, P., 2012. Using a crop model to evaluate and
design combinations of genotypes x management x environments that
improve sunflower crop performance., in: 18th International Sunflower
Conference, 27 February - 1st March, Mar Del Plata, Argentina.

\hypertarget{ref-Casadebaig2011a}{}
Casadebaig, P., Debaeke, P., 2011. Using a crop model to assess
genotype-environment interactions in multi-environment trials, in:
Halford, N., Semenov, M. (Eds.), Aspects of Applied Biology, System
Approaches to Crop Improvement. pp. 19--25.

\hypertarget{ref-Casadebaig2008}{}
Casadebaig, P., Debaeke, P., Lecoeur, J., 2008. Thresholds for leaf
expansion and transpiration response to soil water deficit in a range of
sunflower genotypes. European Journal of Agronomy 28, 646--654.
doi:\href{https://doi.org/10.1016/j.eja.2008.02.001}{10.1016/j.eja.2008.02.001}

\hypertarget{ref-Casadebaig2011}{}
Casadebaig, P., Guilioni, L., Lecoeur, J., Christophe, A., Champolivier,
L., Debaeke, P., 2011. SUNFLO, a model to simulate genotype-specific
performance of the sunflower crop in contrasting environments.
Agricultural and Forest Meteorology 151, 163--178.
doi:\href{https://doi.org/10.1016/j.agrformet.2010.09.012}{10.1016/j.agrformet.2010.09.012}

\hypertarget{ref-Casadebaig2014}{}
Casadebaig, P., Trépos, R., Picheny, V., Langlade, N.B., Vincourt, P.,
Debaeke, P., 2014. Increased genetic diversity improves crop yield
stability under climate variability: A computational study on sunflower.
arXiv preprint arXiv:1403.2825.

\hypertarget{ref-Champolivier2011}{}
Champolivier, L., Debaeke, P., Thibierge, J., Dejoux, J.-F., Ledoux, S.,
Ludot, M., Berger, F., Casadebaig, P., Jouffret, P., Vogrincic, C.,
Lecomte, V., Merrien, A., Mestries, E., Thiard, J., Mistou, M.-N.,
Caumes, E., Edeline, T., Provot, M., 2011. Construire des stratégies de
production adaptées aux débouchés à l'échelle du bassin de collecte.
Innovations Agronomiques 14, 39--57.

\hypertarget{ref-Chapman2008}{}
Chapman, S., 2008. Use of crop models to understand genotype by
environment interactions for drought in real-world and simulated plant
breeding trials. Euphytica 161, 195--208.
doi:\href{https://doi.org/10.1007/s10681-007-9623-z}{10.1007/s10681-007-9623-z}

\hypertarget{ref-Chapman2002}{}
Chapman, S., Cooper, M., Hammer, G., 2002. Using crop simulation to
generate genotype by environment interaction effects for sorghum in
water-limited environments. Australian Journal of Agricultural Research
53, 379--389.

\hypertarget{ref-Chenu2011}{}
Chenu, K., Cooper, M., Hammer, G., Mathews, K., Dreccer, M., Chapman,
S., 2011. Environment characterization as an aid to wheat improvement:
Interpreting genotype--environment interactions by modelling
water-deficit patterns in North-Eastern Australia. Journal of
Experimental Botany 62, 1743--1755.
doi:\href{https://doi.org/10.1093/jxb/erq459}{10.1093/jxb/erq459}

\hypertarget{ref-Cohen1968}{}
Cohen, J., 1968. Weighted kappa: Nominal scale agreement provision for
scaled disagreement or partial credit. Psychological bulletin 70, 213.
doi:\href{https://doi.org/10.1037/h0026256}{10.1037/h0026256}

\hypertarget{ref-CTPS2014}{}
CTPS, 2014. Règlement technique d'examen des variétés de tournesol en
vue de leur inscription au catalogue officiel français (liste a et liste
b). CTPS.

\hypertarget{ref-Vega2012}{}
de la Vega, A., 2012. Effect of the complexity of sunflower growing
regions on the genetic progress achieved by breeding programs. Helia 35,
113--122.
doi:\href{https://doi.org/10.2298/hel1257113v}{10.2298/hel1257113v}

\hypertarget{ref-Vega2001a}{}
de la Vega, A.J., Chapman, S.C., Hall, A.J., 2001. Genotype by
environment interaction and indirect selection for yield in sunflower:
I. Two-mode pattern analysis of oil and biomass yield across
environments in Argentina. Field Crops Research 72, 17--38.
doi:\href{https://doi.org/10.1016/s0378-4290(01)00162-9}{10.1016/s0378-4290(01)00162-9}

\hypertarget{ref-Debaeke2012b}{}
Debaeke, P., Barbet-Massin, C., Salvi, F., Uyttewaal, V., 2012. A
model-based evaluation of the representativeness of multi-environment
trials used for sunflower variety assessment in france., in: 12th Esa
Congress, Helsinki, Finland. pp. 322--323.

\hypertarget{ref-Debaeke2010}{}
Debaeke, P., Casadebaig, P., Haquin, B., Mestries, E., Palleau, J.-P.,
Salvi, F., 2010. Simulation de la réponse variétale du tournesol à
l'environnement à l'aide du modèle sunflo. Oilseeds and fats, Crops and
Lipids 17, 143--51.
doi:\href{https://doi.org/10.1684/ocl.2010.0308}{10.1684/ocl.2010.0308}

\hypertarget{ref-Debaeke2011a}{}
Debaeke, P., Casadebaig, P., Mestries, E., Palleau, J.-P., Salvi, F.,
Bertoux, V., Uyttewaal, V., 2011. Evaluer et valoriser les interactions
variété-milieu-conduite en tournesol. Innovations Agronomiques 14,
77--90.

\hypertarget{ref-DeLacy1996}{}
DeLacy, I., Basford, K., Cooper, M., Bull, J., McLaren, C., 1996.
Analysis of multi-environment trials--an historical perspective. Plant
adaptation and crop improvement 39124.

\hypertarget{ref-Foucteau2001}{}
Foucteau, V., El Daouk, M., Baril, C., 2001. Interpretation of genotype
by environment interaction in two sunflower experimental networks. TAG
Theoretical and Applied Genetics 102, 327--334.
doi:\href{https://doi.org/10.1007/s001220051649}{10.1007/s001220051649}

\hypertarget{ref-GarciaLopez2014}{}
García-López, J., Lorite, I.J., García-Ruiz, R., Domínguez, J., 2014.
Evaluation of three simulation approaches for assessing yield of rainfed
sunflower in a mediterranean environment for climate change impact
modelling. Climatic change 124, 147--162.
doi:\href{https://doi.org/10.1007/s10584-014-1067-6}{10.1007/s10584-014-1067-6}

\hypertarget{ref-Granier1998}{}
Granier, C., Tardieu, F., 1998. Is thermal time adequate for expressing
the effects of temperature on sunflower leaf development? Plant, Cell \&
Environment 21, 695--703.
doi:\href{https://doi.org/10.1046/j.1365-3040.1998.00319.x}{10.1046/j.1365-3040.1998.00319.x}

\hypertarget{ref-Hiederer2013}{}
Hiederer, R., 2013. Mapping soil properties for europe: Spatial
representation of soil database attributes. JRC, Luxembourg:
Publications Office of the European Union, EUR26082EN Scientific;
Technical Research series, ISSN 1831-9424; Citeseer.
doi:\href{https://doi.org/10.2788/94128}{10.2788/94128}

\hypertarget{ref-Holzworth2014}{}
Holzworth, D.P., Huth, N.I., deVoil, P.G., Zurcher, E.J., Herrmann,
N.I., McLean, G., Chenu, K., Oosterom, E.J. van, Snow, V., Murphy, C.,
Moore, A.D., Brown, H., Whish, J.P., Verrall, S., Fainges, J., Bell,
L.W., Peake, A.S., Poulton, P.L., Hochman, Z., Thorburn, P.J., Gaydon,
D.S., Dalgliesh, N.P., Rodriguez, D., Cox, H., Chapman, S., Doherty, A.,
Teixeira, E., Sharp, J., Cichota, R., Vogeler, I., Li, F.Y., Wang, E.,
Hammer, G.L., Robertson, M.J., Dimes, J.P., Whitbread, A.M., Hunt, J.,
Rees, H. van, McClelland, T., Carberry, P.S., Hargreaves, J.N., MacLeod,
N., McDonald, C., Harsdorf, J., Wedgwood, S., Keating, B.A., 2014. APSIM
- Evolution towards a new generation of agricultural systems simulation.
Environmental Modelling \& Software 62, 327--350.
doi:\href{https://doi.org/http://dx.doi.org/10.1016/j.envsoft.2014.07.009}{http://dx.doi.org/10.1016/j.envsoft.2014.07.009}

\hypertarget{ref-Jeuffroy2014}{}
Jeuffroy, M.-H., Casadebaig, P., Debaeke, P., Loyce, C., Meynard, J.-M.,
2014. Agronomic model uses to predict cultivar performance in various
environments and cropping systems. a review. Agronomy for Sustainable
Development 34, 121--137.
doi:\href{https://doi.org/10.1007/s13593-013-0170-9}{10.1007/s13593-013-0170-9}

\hypertarget{ref-Jones2003}{}
Jones, J.W., Hoogenboom, G., Porter, C., Boote, K., Batchelor, W., Hunt,
L., Wilkens, P., Singh, U., Gijsman, A., Ritchie, J., 2003. The dssat
cropping system model. European journal of agronomy 18, 235--265.
doi:\href{https://doi.org/10.1016/s1161-0301(02)00107-7}{10.1016/s1161-0301(02)00107-7}

\hypertarget{ref-Kendall1948}{}
Kendall, M.G., 1948. Rank correlation methods. 160 p.

\hypertarget{ref-Landis1977}{}
Landis, J.R., Koch, G.G., 1977. The measurement of observer agreement
for categorical data. biometrics 159--174.
doi:\href{https://doi.org/10.2307/2529310}{10.2307/2529310}

\hypertarget{ref-Lecoeur2011}{}
Lecoeur, J., Poiré-Lassus, R., Christophe, A., Pallas, B., Casadebaig,
P., Debaeke, P., Vear, F., Guilioni, L., 2011. Quantifying physiological
determinants of genetic variation for yield potential in sunflower.
SUNFLO: a model-based analysis. Functional Plant Biology 38, 246--259.
doi:\href{https://doi.org/10.1071/fp09189}{10.1071/fp09189}

\hypertarget{ref-Lecoeur1996a}{}
Lecoeur, J., Sinclair, T.R., 1996. Field Pea Transpiration and Leaf
Growth in Response to Soil Water Deficit. Crop Science 36, 331--335.
doi:\href{https://doi.org/doi:10.2135/cropsci1996.0011183X003600020020x}{doi:10.2135/cropsci1996.0011183X003600020020x}

\hypertarget{ref-Lecomte2010}{}
Lecomte, C., Prost, L., Cerf, M., Meynard, J., 2010. Basis for designing
a tool to evaluate new cultivars. Agronomy for sustainable development
30, 667--677.
doi:\href{https://doi.org/10.1051/agro/2009042}{10.1051/agro/2009042}

\hypertarget{ref-Leite2014}{}
Leite, J.G.D.B., Silva, J.V., Justino, F.B., Ittersum, M.K. van, 2014. A
crop model-based approach for sunflower yields. Scientia agricola 71,
345--355.
doi:\href{https://doi.org/10.1590/0103-9016-2013-0356}{10.1590/0103-9016-2013-0356}

\hypertarget{ref-Malosetti2013}{}
Malosetti, M., Ribaut, J.-M., Eeuwijk, F.A. van, 2013. The statistical
analysis of multi-environment data: Modeling genotype-by-environment
interaction and its genetic basis. Frontiers in physiology 4.
doi:\href{https://doi.org/10.3389/fphys.2013.00044}{10.3389/fphys.2013.00044}

\hypertarget{ref-Marinkovic2011}{}
Marinković, R., Jocković, M., Marjanović-Jeromela, A., Jocić, S., Ćirić,
M., Balalić, I., Sakač, Z., 2011. Genotype by environment interactions
for seed yield and oil content in sunflower (h. annuus l.) using AMMI
model. Helia 34, 79--88.
doi:\href{https://doi.org/10.2298/hel1154079m}{10.2298/hel1154079m}

\hypertarget{ref-Martre2015a}{}
Martre, P., Wallach, D., Asseng, S., Ewert, F., Jones, J.W., Rötter,
R.P., Boote, K.J., Ruane, A.C., Thorburn, P.J., Cammarano, D., Hatfield,
J.L., Rosenzweig, C., Aggarwal, P.K., Angulo, C., Basso, B., Bertuzzi,
P., Biernath, C., Brisson, N., Challinor, A.J., Doltra, J., Gayler, S.,
Goldberg, R., Grant, R.F., Heng, L., Hooker, J., Hunt, L.A., Ingwersen,
J., Izaurralde, R.C., Kersebaum, K.C., Müller, C., Kumar, S.N., Nendel,
C., O'leary, G., Olesen, J.E., Osborne, T.M., Palosuo, T., Priesack, E.,
Ripoche, D., Semenov, M.A., Shcherbak, I., Steduto, P., Stöckle, C.O.,
Stratonovitch, P., Streck, T., Supit, I., Tao, F., Travasso, M., Waha,
K., White, J.W., Wolf, J., 2015. Multimodel ensembles of wheat growth:
Many models are better than one. Global Change Biology 21, 911--925.
doi:\href{https://doi.org/10.1111/gcb.12768}{10.1111/gcb.12768}

\hypertarget{ref-Merrien1992}{}
Merrien, A., 1992. Les points techniques du CETIOM : Physiologie du
tournesol. CETIOM.

\hypertarget{ref-Mestries2002}{}
Mestries, E., Jouffret, P., 2002. Comment le CETIOM évalue les variétés.
Oléoscope 66, 4--8.

\hypertarget{ref-Monteith1994}{}
Monteith, J.L., 1994. Validity of the correlation between intercepted
radiation and biomass. Agricultural and Forest Meteorology 68, 213--220.
doi:\href{https://doi.org/10.1016/0168-1923(94)90037-x}{10.1016/0168-1923(94)90037-x}

\hypertarget{ref-Monteith1977}{}
Monteith, J.L., 1977. Climate and the Efficiency of Crop Production in
Britain. Philosophical Transactions of the Royal Society of London.
Series B, Biological Sciences 281, 277--294.
doi:\href{https://doi.org/10.2307/2402584}{10.2307/2402584}

\hypertarget{ref-Pidgeon2006}{}
Pidgeon, J.D., Ober, E.S., Qi, A., Clark, C.J., Royal, A., Jaggard,
K.W., 2006. Using multi-environment sugar beet variety trials to screen
for drought tolerance. Field crops research 95, 268--279.
doi:\href{https://doi.org/10.1016/j.fcr.2005.04.010}{10.1016/j.fcr.2005.04.010}

\hypertarget{ref-Piepho2012}{}
Piepho, H.-P., Möhring, J., Schulz-Streeck, T., Ogutu, J.O., 2012. A
stage-wise approach for the analysis of multi-environment trials.
Biometrical Journal 54, 844--860.
doi:\href{https://doi.org/10.1002/bimj.201100219}{10.1002/bimj.201100219}

\hypertarget{ref-Quere2004}{}
Quere, L., 2004. Des facteurs clés limitants pour le tournesol
identifiés en 2003. Oléoscope 31--32.

\hypertarget{ref-Rey2008}{}
Rey, H., Dauzat, J., Chenu, K., Barczi, J.-F., Dosio, G.A.A., Lecoeur,
J., 2008. Using a 3-D Virtual Sunflower to Simulate Light Capture at
Organ, Plant and Plot Levels: Contribution of Organ Interception, Impact
of Heliotropism and Analysis of Genotypic Differences. Annals of Botany
1139--1152.
doi:\href{https://doi.org/10.1093/aob/mcm300}{10.1093/aob/mcm300}

\hypertarget{ref-Robertson2015}{}
Robertson, M., Rebetzke, G., Norton, R., 2015. Assessing the place and
role of crop simulation modelling in australia. Crop and Pasture
Science.

\hypertarget{ref-Rosenzweig2013}{}
Rosenzweig, C., Jones, J., Hatfield, J., Ruane, A., Boote, K., Thorburn,
P., Antle, J., Nelson, G., Porter, C., Janssen, S., Asseng, S., Basso,
B., Ewert, F., Wallach, D., Baigorria, G., Winter, J., 2013. The
agricultural model intercomparison and improvement project (agmip):
Protocols and pilot studies. Agricultural and Forest Meteorology 170,
166--182.
doi:\href{https://doi.org/10.1016/j.agrformet.2012.09.011}{10.1016/j.agrformet.2012.09.011}

\hypertarget{ref-Rouse1988}{}
Rouse, D., 1988. Use of crop growth-models to predict the effects of
disease. Annual review of Phytopathology 26, 183--201.
doi:\href{https://doi.org/10.1146/annurev.phyto.26.1.183}{10.1146/annurev.phyto.26.1.183}

\hypertarget{ref-Savary2000}{}
Savary, S., Willocquet, L., Elazegui, F., Castilla, N., Teng, P., 2000.
Rice pest constraints in tropical asia: Quantification of yield losses
due to rice pests in a range of production situations. Plant disease 84,
357--369.
doi:\href{https://doi.org/10.1094/pdis.2000.84.3.357}{10.1094/pdis.2000.84.3.357}

\hypertarget{ref-Sinclair1986a}{}
Sinclair, T., Ludlow, M., 1986. Influence of Soil Water Supply on the
Plant Water Balance of Four Tropical Grain Legumes. Australian Journal
of Plant Physiology 13, 329--341.

\hypertarget{ref-Sinclair2001}{}
Sinclair, T.R., Muchow, R.C., 2001. System Analysis of Plant Traits to
Increase Grain Yield on Limited Water Supplies. Agronomy Journal 93,
263--270.

\hypertarget{ref-TerresInovia2016}{}
Terres Inovia, 2016. Guide de culture tournesol. Terres OléoPro.

\hypertarget{ref-Triboi2004}{}
Triboi, A.M., Messaoud, J., Debaeke, P., Lecoeur, J., Vear, F., 2004.
Heredity of sunflower leaf characters useable as yield predictors., in:
Seiler, G. (Ed.), Proceedings of the 16th International Sunflower
Conference. Fargo, North Dakota, USA., pp. 517--523.

\hypertarget{ref-Vale2007}{}
Valé, M., Mary, B., Justes, E., 2007. Irrigation practices may affect
denitrification more than nitrogen mineralization in warm climatic
conditions. Biology and Fertility of Soils 43, 641--651.
doi:\href{https://doi.org/10.1007/s00374-006-0143-0}{10.1007/s00374-006-0143-0}

\hypertarget{ref-VanWaes2009}{}
Van Waes, J., 2009. Maize variety testing for registration on a national
catalogue and the impact of new technologies. Maydica 54, 139.

\hypertarget{ref-Verger2014}{}
Verger, A., Vigneau, N., Chéron, C., Gilliot, J.-M., Comar, A., Baret,
F., 2014. Green area index from an unmanned aerial system over wheat and
rapeseed crops. Remote Sensing of Environment 152, 654--664.
doi:\href{https://doi.org/10.1016/j.rse.2014.06.006}{10.1016/j.rse.2014.06.006}

\hypertarget{ref-Wallach2014}{}
Wallach, D., Makowski, D., Jones, J.W., Brun, F., 2014. Working with
dynamic crop models. methods, tools and examples for agriculture and
environment, 2nd ed. Academic Press.

\hypertarget{ref-Welham2010}{}
Welham, S.J., Gogel, B.J., Smith, A.B., Thompson, R., Cullis, B.R.,
2010. A comparison of analysis methods for late-stage variety evaluation
trials. Australian \& New Zealand Journal of Statistics 52, 125--149.
doi:\href{https://doi.org/10.1111/j.1467-842x.2010.00570.x}{10.1111/j.1467-842x.2010.00570.x}

\hypertarget{ref-Willocquet2008}{}
Willocquet, L., Aubertot, J., Lebard, S., Robert, C., Lannou, C.,
Savary, S., 2008. Simulating multiple pest damage in varying winter
wheat production situations. Field Crops Research 107, 12--28.
doi:\href{https://doi.org/10.1016/j.fcr.2007.12.013}{10.1016/j.fcr.2007.12.013}

\end{document}